\documentclass[aps,prl]{revtex4}
\usepackage{amsmath,amssymb}
\usepackage{color}
\usepackage[pdftex]{graphicx}
\usepackage{xcolor}
\usepackage[makeroom]{cancel}
\usepackage{appendix}
\usepackage{soul}
\usepackage[shortlabels]{enumitem}
\usepackage{caption}
\DeclareMathSymbol{\R}{\mathalpha}{AMSb}{"52}

\usepackage[a4paper,margin=1in]{geometry}
\usepackage{fancyhdr}

\DeclareMathAlphabet\mathbfcal{OMS}{cmsy}{b}{n}

\begin{document}
\title{
Transport Model for the Propagation of Partially-Coherent, \\Polarization-Gradient Vector Beams}
\author{J. M. Nichols}
\affiliation{Naval Research Laboratory\\ 4555 Overlook Ave. SW.\\ Washington D.C. 20375}
\author{D. V. Nickel} 
\affiliation{Naval Research Laboratory\\ 4555 Overlook Ave. SW.\\ Washington D.C. 20375}
\author{F. Bucholtz}
\affiliation{Jacobs Technology, Inc.\\ 2551 Dulles View Drive\\ Herndon, VA 20171}
\author{G. Rohde }
\affiliation{University of Virginia\\ Dept. of Electrical and Computer Engineering\\ Dept. of Biomedical Engineering\\
415 Lane Rd.\\Charlottesville, VA}

\begin{abstract}
In a recent work \cite{Nichols:22}, we predicted and experimentally validated a new physical mechanism for altering the propagation path of a monochromatic beam.  Specifically, we showed that by properly tailoring the spatial distribution of the linear state of polarization transverse to the direction of propagation, the beam followed a curved trajectory in free space.  Here we extend the model to the partially coherent, polychromatic case by redefining the beam amplitude, phase, and polarization angle as appropriate statistical quantities.  In particular, we propose an entirely new definition of linear polarization gradient as an average over the third generalized Stokes parameter in the spatial frequency domain.  In the new model, the beam curvature matches that of our previous work in the fully coherent case, but is predicted to gradually vanish as the beam loses coherence and becomes depolarized. The model also clearly predicts, however, that there does not exist a natural mechanism in free space or due to atmospheric turbulence that will cause significant depolarization. Simulated beam trajectories are shown for varying levels of initial partial coherence and for different polarization profiles. A new class of non-diffracting beams is also suggested by way of example.  Lastly, as a byproduct of the derivation we 
generalize a previously known result concerning the free-space propagation of generalized Stokes parameters \cite{Korotkova:08} by showing the result also holds for propagation in inhomogeneous media.   
\end{abstract}

\maketitle

\section{Introduction}
Recently we demonstrated that through proper choice of spatial distribution of linear polarization angle, a coherent, monochromatic beam follows a curved trajectory in free space  \cite{Nichols:22}.  The effect is similar in magnitude to that of Airy beam bending (both scale as $k_0^{-2}$ where $k_0$ is the free-space wavenumber).  However, in our approach it is the centroid of the beam that experiences a transverse acceleration as opposed to the Airy beam case where the main intensity lobe transversely accelerates, but where the centroid of the beam does not.  The physics of the vector beam bending were predicted via a transport model, whereby the parameters defining the optical field (intensity and phase) were shown to be governed by the transport of intensity equation (TIE) and an eikonal equation, expressing conservation of transverse linear momentum.  The model was re-formulated in Lagrangian coordinates in order to solve for the transverse beam location which, under appropriate choice of polarization gradient, was shown to bend in the transverse plane.  These predictions were subsequently verified in experiment \cite{Nichols:22}.
\pagestyle{plain} % revert to "no footer" page style

In this work we extend the model to the more general case of partially coherent light, showing that with appropriate definitions of optical phase, amplitude, and polarization gradient the model still predicts a bending effect. For perfectly coherent light, the governing equations reduce to the previously derived coherent case, while for incoherent light, the effect disappears.  The resulting model also predicts that the state of linear polarization (including the degree of polarization) will remain unchanged on propagation. 
 Collectively, these model predictions represent the main contribution of this work.  

A second, but related contribution is to provide a new definition of polarization angle gradient that is appropriate for modeling partially coherent vector beams.  Our definition is based on the normalized integral of the third generalized Stokes parameter and incorporates the degree of polarization directly.  While this definition falls naturally from the vector transport equations, it is also consistent with recent work on scalar, partially coherent beams where a similar definition of generalized phase was proposed \cite{Zuo:15}.  In what follows, we show that this scalar definition is also easily extended to the vector case and is also a natural consequence of the transport model.

Finally, in the process of deriving this result, we generalize a previously known result from Korotkova \cite{Korotkova:08}.  In that work, the author showed that the transverse integral of the generalized Stokes parameters were conserved under free-space propagation.  We confirm this result using an entirely different approach and extend it to the case of propagation in an inhomogeneous medium.

Our approach is as follows. \\[-14pt]
\begin{enumerate}
\item We begin with the wave equation in the general case of a weakly-inhomogenous medium. We make the paraxial assumption via the slowly-varying envelope approximation to obtain Helmholtz equations for each field component.  In the instance where the medium warrants a probabilistic description, we additionally invoke the Markov Approximation \cite{Tatarskii:69}. \\[-14pt]
\item To accommodate the probabilistic nature of the partially coherent problem, the fields are represented in the coherency matrix formalism where each matrix element is an expectation of a product of finite-time Fourier transforms \cite{Priestly:81} of field components at two different transverse spatial locations. Importantly, we retain the dependence on spatially-separated points throughout the calculation in the form of Wigner Transforms with respect to the transverse spatial coordinate.  This information is vital in our model as the state of polarization is not uniform in the transverse plane. \\[-14pt] 
\item We obtain a vector transport equation for the generalized Stokes parameters which is the partially coherent, vector counterpart to the transport of intensity equations obtained in our earlier paper \cite{Nichols:22}. It is entirely reasonable to expect that, in the case of partial polarization, individual components of the Stokes 4-vector must obey continuity equations while, in the fully-polarized case, continuity of the intensity is sufficient to characterize beam propagation.
\item We extend the method of moments developed for the scalar problem \cite{Marcuvitz:80,Degond:03} to the vector case, leading to the coupled transport equations describing the evolution of expected amplitude, phase, and polarization angle. Importantly, the model reduces to our monochromatic, coherent model in those respective limits.
\end{enumerate}
The resulting model extends our prior coherent model variables (phase, polarization angle, and intensity) to the partially coherent case by taking appropriate averages over the Wigner spatial frequency and with these new definitions, predict the {\it expected} beam path.

\section{Transport Modeling of Partially Coherent, Polarized Light \label{sec:transport}}

\subsection{Spectral Coherency Representation}

We start with the vector wave equation for electric-field vector $\mathbf{E}({\bf x},t)$
\begin{align}
\nabla^2{\bf E}({\bf x},t)-\frac{\epsilon({\bf x})}{c_0^2}\frac{\partial^2{\bf E}({\bf x},t)}{\partial t^2}&=0
\label{eqn:waveEqn}
\end{align}
 which presumes an inhomogeneous, isotropic medium characterized by dimensionless dielectric constant $\epsilon({\bf x})>1$, where $\mathbf{x}$ denotes position in three-dimensional Cartesian space, where $c_0=1/\sqrt{\epsilon_0\mu_0}$ is the speed of light in vacuum, and where $\epsilon_0$ and $\mu_0$ are the permittivity and permeability, respectively, of free space. We assume the inhomogeneity is weak and define 
 \begin{align}
 \epsilon_1({\bf x})=\dfrac{\epsilon({\bf x})-\left\langle\epsilon\right\rangle}{\left\langle\epsilon\right\rangle}<<1
 \end{align}
 where $\left\langle\epsilon\right\rangle$ is the average value of $\epsilon({\bf x})$ taken over the entire region of interest.  
 
 We assume the wave propagates in the $+z$-direction and we model the electric field vector as a stationary random process, existing for times $-T\le t\le T$.  Since (\ref{eqn:waveEqn}) is linear in ${\bf E}({\bf x},t)$, we may consider a superposition of monochromatic, transverse-wave solutions
\begin{align}
    {\bf E}({\bf x},t)&=\frac{1}{2\pi}\int_{-\infty}^{\infty}\left\{E_X({\bf x},\omega)_T,E_Y({\bf x},\omega)_T,0\right\}e^{-ik_0z}e^{i\omega t}d\omega,\hspace{0.8cm}-T\le t\le T
    \label{eqn:planewave}
\end{align}
represented here by the Fourier integral. Phase accumulation in the direction of propagation is captured by the $e^{-ik_0 z}$ term with wavenumber  $k_0=(\omega/c_0)\left\langle\epsilon\right\rangle^{1/2}=(2\pi/\lambda)\left\langle\epsilon\right\rangle^{1/2}$ at free-space wavelength $\lambda$. The quantities $\{E_X({\bf x},\omega)_T,~E_Y({\bf x},\omega)_T,0\}e^{-ik_0 z}$ are therefore the associated Fourier amplitudes at frequency $\omega$ and possess units of electric field per Hz.  Given the above construction, these are given by 
\begin{align}
e^{-ik_0 z}E_{X,Y}({\bf x},\omega)_T&=\int_{-T}^{T}E_{X,Y}({\bf x},t)e^{-i\omega t}dt
\end{align}
as discussed in \cite{Priestly:81} (sec. 4.7).  The explicit dependence of the Fourier amplitudes on $T$ will be retained for the moment.  The electric field and its Fourier amplitudes are distinguished by their arguments ($t$ and $\omega$ resp.).  

Substitute (\ref{eqn:planewave}) into (\ref{eqn:waveEqn}) and make the slowly-varying envelope approximation in which it is assumed that changes in the amplitude of the field over $z-$distances of a wavelength are small compared to the amplitude itself. Then $|\partial_{zz}(\cdot)|\ll |k_0\partial_z(\cdot)|$ and the net result is that spatial gradients in ${\bf x}$ become gradients only in the transverse plane leading to the parabolic equations
\begin{align}
\left[
-i2k_0 \partial_z +\nabla^2_X+k_0^2\epsilon_1(\vec{x},z)
\right]
\left\{E_X(\vec{x},z,\omega)_T,E_Y(\vec{x},z,\omega)_T\right\}
=0\nonumber \\
\left[
i2k_0 \partial_z +\nabla^2_X+k_0^2\epsilon_1(\vec{x},z)
\right]
\left\{E_X^*(\vec{x},z,\omega)_T,E_Y^*(\vec{x},z,\omega)_T\right\}
=0
\label{eqn:paraxial1}
\end{align}
which must be satisfied by each vector component of Fourier amplitudes and their complex conjugates (denoted with $^*$).  Here we have denoted the position in the transverse plane as the vector $\vec{x}\equiv \{x,y\}$ to distinguish it from the propagation direction $z$ (${\bf x}\equiv \{\vec{x},z\}$), and have denoted the transverse gradient operator $\nabla_X(\cdot)\equiv \{\partial_x(\cdot),~\partial_y(\cdot)\}$.  

Note that (\ref{eqn:paraxial1}) comprises four separate equations in, respectively,  $E_X,E_Y,E_X^*,E_Y^*$ which we will call (a-d).  Now perform the following operations.
Multiply (a) and (b) evaluated at $\vec{x}_1$ by $E_X^*(\vec{x}_2,z,\omega)$.  Multiply (a) and (b) evaluated at $\vec{x}_1$ by $E_Y^*(\vec{x}_2,z,\omega)$.  Multiply (c) and (d) evaluated at $\vec{x}_2$ by $E_X(\vec{x}_1,z,\omega)$.  Multiply (c) and (d) evaluated at $\vec{x}_2$ by $E_Y(\vec{x}_1,z,\omega)$.  Note that in each of the above operations all four multiplications are performed from the right.  The results, in order, are eight equations which we denote (i)-(viii).

If we then subtract (v) from (i), (vii) from (ii), (vi) from (iii), and (viii) from (iv) we find (taking into account the chain rule for the derivatives in $z$):
\begin{align}
\left[
i2k_0\partial_z
+\left(\nabla^2_{X_2}-\nabla^2_{X_1}\right)+
k_0^2\left(\epsilon_1(\vec{x}_2,z)-\epsilon_1(\vec{x}_1,z)\right)
\right]
E_i(\vec{x}_1,z,\omega)_T E_j^*(\vec{x}_2,z,\omega)_T=&0,~i,j\in \{X,Y\}
\label{eqn:paraxial2b}
\end{align}
where the notation $\nabla^2_{X_l}$ indicates that the Laplacian operation is to be performed as a function of the transverse variable designated $\vec{x}_l$ for $l=1,2$. Taking the expected value of (\ref{eqn:paraxial2b}) over realizations of the electric fields and performing the limiting operation
 \begin{align}
\left\langle E_i(\vec{x},z,\omega)E^*_j(\vec{x},z,\omega)\right\rangle \equiv \lim_{T\rightarrow\infty}\left(\mathbb{E}\left[\frac{E_i(\vec{x},z,\omega)_T E^*_j(\vec{x},z,\omega)_T }{2T}\right]\right),~i,j\in \{X,Y\}
 \end{align}
 then yields
\begin{align}
    i2 k\partial_z W_{ij}(\vec{x}_1,\vec{x}_2,z,\omega)
    &+\left(\nabla^2_{X_2}-\nabla^2_{X_1}\right)W_{ij}(\vec{x}_1,\vec{x}_2,z,\omega)
    \nonumber \\&+k_0^2\left[\epsilon_1(\vec{x}_2,z)-\epsilon_1(\vec{x}_1,z)\right]W_{ij}(\vec{x}_1,\vec{x}_2,z,\omega)=0,~i,j\in \{X,Y\}
    \label{eqn:paraxial3}
\end{align}
where each $W_{ij}(\vec{x}_1,\vec{x}_2,z,\omega)$ is an element of the spectral density matrix
\begin{align}
    W_{ij}\in {\bf W}(\vec{x}_1,\vec{x}_2,z,\omega)&=\left[\begin{array}{cc}
    \left\langle E_X(\vec{x}_1,z,\omega)E_X^*(\vec{x}_2,z,\omega)\right\rangle & \left\langle E_X(\vec{x}_1,z,\omega)E_Y^*(\vec{x}_2,z,\omega)\right\rangle\\
    \left\langle E_Y(\vec{x}_1,z,\omega)E_X^*(\vec{x}_2,z,\omega)\right\rangle &
    \left\langle E_Y(\vec{x}_1,z,\omega)E_Y^*(\vec{x}_2,z,\omega)\right\rangle 
    \end{array}\right]
    \label{eqn:specohere}
\end{align}
and possesses units of electric field squared per Hz.  Note that multiplying (\ref{eqn:specohere}) by $\epsilon_0/2$  gives ${\bf W}(\vec{x}_1,\vec{x}_2,z,\omega)$ in units of power spectral density in Watts/Hz.  

Equation (\ref{eqn:paraxial3}) treats the medium properties as deterministic, such as propagation through a medium of known refractive index profile, in which case it is common to set $\langle\epsilon\rangle=1$ and $\epsilon_1(\vec{x},z)=n^2(\vec{x},z)-1$ where $n(\vec{x},z)$ is the refractive index.  Alternatively, we can assume the medium is described by its statistical properties.  Treating $\epsilon_1(\vec{x},z)$ as a zero-mean, Gaussian random variable (e.g., as in a turbulent atmosphere) requires averaging over both the field amplitudes {\it and} the medium properties in taking the expectation of Eq. (\ref{eqn:paraxial2b}).  Define these properties via the covariance $\langle \epsilon_1(\vec{x}_2,z)\epsilon_1(\vec{x}_1,z')\rangle=\int_{\R^3} S_{NN}(\vec{\xi},\kappa)e^{i\vec{\xi}\cdot(\vec{x}_2-\vec{x}_1)+i\kappa(z-z')}d\vec{\xi}d\kappa$ , that is, as the inverse Fourier Transform of the three-dimensional spectral density associated with fluctuations in the dielectric constant.  We then invoke the well-established Markov Approximation (MA) of Tatarskii \cite{Tatarskii:69},~\cite{Fante:75} (see extension to the vector electric field case in \cite{Charnotskii:16}) which assumes the medium is ``delta'' correlated in the direction of propagation so that $\langle \epsilon_1(\vec{x}_2,z)\epsilon_1(\vec{x}_1,z')\rangle\approx\delta(z-z')A(\vec{x}_2-\vec{x}_1,z')$.  Under the MA, Tatarskii \cite{Tatarskii:69} showed that the required average in (\ref{eqn:paraxial2b}) can be written
\begin{align}
\bigg\langle\left[\epsilon_1(\vec{x}_2,z)-\epsilon_1(\vec{x}_1,z)\right]
E_i(\vec{x}_1,z,\omega)_T E_j^*(\vec{x}_2,z,\omega)_T\bigg\rangle\approx \frac{ik_0}{2}\left[A(0,z)-A(\vec{x}_2-\vec{x}_1,z)\right]W_{ij}(\vec{x}_1,\vec{x}_2,z,\omega)
\label{eqn:averages}
\end{align}
so that Eq. (\ref{eqn:paraxial3}) becomes
\begin{align}
    i2 k\partial_z W_{ij}(\vec{x}_1,\vec{x}_2,z,\omega)
    &+\left(\nabla^2_{X_2}-\nabla^2_{X_1}\right)W_{ij}(\vec{x}_1,\vec{x}_2,z,\omega)\nonumber \\
    &+\frac{ik_0^3}{2}\left[A(0,z)-A(\vec{x}_2-\vec{x}_1,z)\right]W_{ij}(\vec{x}_1,\vec{x}_2,z,\omega)=0
    \label{eqn:paraxial3b}
\end{align}
As we will show, the  choice of either a deterministic (\ref{eqn:paraxial3}) or a stochastic (\ref{eqn:paraxial3b}) medium can be easily accommodated in the transport model.  Also note that the last term in both (\ref{eqn:paraxial3}) and (\ref{eqn:paraxial3b}) possess units of $[m]$ so that $k A(\cdot)$ is dimensionless, as is $\epsilon(\cdot)$.  Note that in \cite{Charnotskii:16}, the approximation $\epsilon_1\approx 2\eta$ is invoked which would have led to the pre-factor in Eq. (\ref{eqn:averages}) being written as $2ik_0$.  

%For stationary processes as assumed here, the spectral density and the correlation functions $\Gamma_{ij}(\vec{x}_1,\vec{x}_2,z,\tau)\equiv \left\langle E_i(\vec{x}_1,z,t)E_j(\vec{x}_2,z,t+\tau)\right\rangle$ are Fourier transform pairs
%\begin{align}
%W_{ij}(\vec{x}_1,\vec{x}_2,z,\omega)&=\int_{-\infty}^{\infty}\Gamma_{ij}(\vec{x}_1,\vec{x}_2,z,\tau)e^{-i\omega\tau}d\tau,~i,j\in X,Y
%\end{align}
%under the same limiting operation $T\rightarrow\infty$ as was used in defining the spectral density matrix (see e.g., \cite{Priestly:81}). 
Equation (\ref{eqn:paraxial3}) or (\ref{eqn:paraxial3b}) therefore contain four separate equations, each governing a different element of the spectral density matrix (see also Wolf {\it et al.} e.g. \cite{Wolf:05}).  Note that at the spatial location $\vec{x}_1=\vec{x}_2\equiv \vec{x}$ the expressions $\langle E_i(\vec{x}_1,z,\omega)E_j^*(\vec{x}_2,z,\omega)\rangle$ reduce to the more familiar power spectral density functions
\begin{align}
S_{ij}(\vec{x},z,\omega)&=\left\langle E_i(\vec{x},z,\omega)E_j^*(\vec{x},z,\omega)\right\rangle,~i,j\in\{X,Y\}
    \label{eqn:PSD}
\end{align}
that is, $S_{ij}(\vec{x},z,\omega)=W_{ij}(\vec{x},\vec{x},z,\omega)$. 

\subsection{Generalize Stokes Parameter Representation}
It will be convenient to describe propagation in terms of {\it generalized} Stokes parameters \cite{Wolf:05b}. Define
\begin{align}
    s_0(\vec{x}_1,\vec{x}_2,z,\omega)&=\left\langle E_X(\vec{x}_1,z,\omega)E_X^*(\vec{x}_2,z,\omega)\right\rangle+\left\langle E_Y(\vec{x}_1,z,\omega)E_Y^*(\vec{x}_2,z,\omega)\right\rangle\nonumber \\
    &=W_{XX}(\vec{x}_1,\vec{x}_2,z,\omega)+W_{YY}(\vec{x}_1,\vec{x}_2,z,\omega)\nonumber \\
    s_1(\vec{x}_1,\vec{x}_2,z,\omega)&=\left\langle E_X(\vec{x}_1,z,\omega)E_X^*(\vec{x}_2,z,\omega)\right\rangle-\left\langle E_Y(\vec{x}_1,z,\omega)E_Y^*(\vec{x}_2,z,\omega)\right\rangle \nonumber \\
    &=W_{XX}(\vec{x}_1,\vec{x}_2,z,\omega)-W_{YY}(\vec{x}_1,\vec{x}_2,z,\omega)\nonumber \\
    s_2(\vec{x}_1,\vec{x}_2,z,\omega)&=2Re\{\left\langle E_X(\vec{x}_1,z,\omega)E_Y^*(\vec{x}_2,z,\omega)\right\rangle\}=\left\langle E_X(\vec{x}_1,z,\omega)E_Y^*(\vec{x}_2,z,\omega)
    \right\rangle+\left\langle E_Y(\vec{x}_1,z,\omega)E_X^*(\vec{x}_2,z,\omega)\right\rangle\nonumber \\
    &=W_{XY}(\vec{x}_1,\vec{x}_2,z,\omega)+W_{YX}(\vec{x}_1,\vec{x}_2,z,\omega)\nonumber \\
    s_3(\vec{x}_1,\vec{x}_2,z,\omega)&=-2Im\{\left\langle E_X(\vec{x}_1,z,\omega)E_Y^*(\vec{x}_2,z,\omega)\right\rangle \}=i\left(\left\langle E_X(\vec{x}_1,z,\omega)E_Y^*(\vec{x}_2,z,\omega)\right\rangle-\left\langle E_Y(\vec{x}_1,z,\omega)E_X^*(\vec{x}_2,z,\omega)\right\rangle \right)\nonumber \\
    &=i\left(W_{XY}(\vec{x}_1,\vec{x}_2,z,\omega)-W_{YX}(\vec{x}_1,\vec{x}_2,z,\omega)\right).
    \label{eqn:genStokes1}
\end{align}

Unlike conventional Stokes parameters which are evaluated at a single point in space, these generalized parameters depend on the field relationships at two points $\vec{x}_1$ and $\vec{x}_2$. As $\vec{x}_2\rightarrow \vec{x}_1\equiv \vec{x}$ we recover the standard definitions.  If the above holds, we can also write
\begin{align}
    {\bf W}(\vec{x}_1,\vec{x}_2,z,\omega)&=\frac{1}{2}\left[\begin{array}{cc}
    s_0(\vec{x}_1,\vec{x}_2,z,\omega)+s_1(\vec{x}_1,\vec{x}_2,z,\omega) & s_2(\vec{x}_1,\vec{x}_2,z,\omega)-is_3(\vec{x}_1,\vec{x}_2,z,\omega)\\
    s_2(\vec{x}_1,\vec{x}_2,z,\omega)+is_3(\vec{x}_1,\vec{x}_2,z,\omega) &
    s_0(\vec{x}_1,\vec{x}_2,z,\omega)-s_1(\vec{x}_1,\vec{x}_2,z,\omega)
    \end{array}\right]
    \label{eqn:specohereAlt}
\end{align}
which matches the result of \cite{Burridge:75} (Eq. 3.54) and \cite{Ryzhik:96} (Eq. 1.5), the only difference being the sign convention chosen for $\exp{(i\omega t)}$.  
 \subsection{Wigner Representation}
Because (\ref{eqn:paraxial3},\ref{eqn:paraxial3b}) are linear in $W_{ij}(\vec{x}_1,\vec{x}_2,z,\omega)$ we can consider still other forms.  For example, (\ref{eqn:paraxial3}) is also seen to govern the generalized Stokes parameters,
%\begin{align}
%    i2k_0\partial_z\left(\begin{array}{c} s_0\\s_1\\s_2\\s_3\end{array}\right)+(\nabla^2_{X_2}-\nabla^2_{X_1})\left(\begin{array}{c} s_0\\s_1\\s_2\\s_3\end{array}\right)+k_0^2\left[n^2(\vec{x}_2)-n^2(\vec{x}_1)\right]\left(\begin{array}{c} s_0\\s_1\\s_2\\s_3\end{array}\right)=\left(\begin{array}{c} 0\\0\\0\\0\end{array}\right)
%\end{align}
\begin{align}
\left[
    i2k_0\partial_z+\left(\nabla^2_{X_2}-\nabla^2_{X_1}\right)+k_0^2\left(\epsilon_1(\vec{x}_2)-\epsilon_1(\vec{x}_1)\right)   \right]s_\nu=0,\,\,\,\,\,\,\,\,\,\,\,\,\,\,\nu=0,\cdots,3
\end{align}
where $s_\nu=s_\nu\left(\vec{x}_1,\vec{x}_2,z,\omega\right)$ 
 and where we have omitted the arguments for brevity.  Applying the coordinate transformation $\vec{x}_{1,2}\rightarrow \vec{x}\mp \vec{x}\,'/2$ and noting that, in this new system the operator $(\nabla^2_{X_2}-\nabla^2_{X_1})\rightarrow 2\nabla_{X}\cdot\nabla_{X'}$ \cite{Petruccelli:13}, we have
%\begin{align}
%    i2k_0\partial_z\left(\begin{array}{c} s_0'\\s_1'\\s_2'\\s_3'\end{array}\right)+2(\nabla_{X}\cdot\nabla_{X'})\left(\begin{array}{c} s_0'\\s_1'\\s_2'\\s_3'\end{array}\right)+2k_0^2\Phi(\vec{x},\vec{x}',z)\left(\begin{array}{c} s_0'\\s_1'\\s_2'\\s_3'\end{array}\right)=\left(\begin{array}{c} 0\\0\\0\\0\end{array}\right).
%    \label{eqn:parabolic5}
%\end{align}
\begin{align}   \left[i2k_0\partial_z+2(\nabla_{X}\cdot\nabla_{X'})+2k_0^2\Phi(\vec{x},\vec{x}\,',z)
    \right]s_\nu '=0,\,\,\,\,\,\,\,\,\,\,\,\,\,\,\nu=0,\cdots,3
    \label{eqn:parabolic5}
\end{align}
where we again use the shorthand notation $s_\nu '= s_\nu'(\vec{x},\vec{x}\,',z,\omega)$
and where
\begin{align}
    \Phi(\vec{x},\vec{x}\,',z)&=\frac{1}{2}\left(\epsilon_1(\vec{x}+\frac{\vec{x}\,'}{2},z)-\epsilon_1(\vec{x}-\frac{\vec{x}\,'}{2},z)\right)
    \label{eqn:vecPot}
\end{align}
will be referred to as the Wigner potential function. In free space $\Phi(\vec{x},\vec{x}',z)=0$ while in the case of a stochastic medium, the Wigner potential of Eq. (\ref{eqn:parabolic5}) becomes 
\begin{align}
    \Phi(\vec{x},\vec{x}',z)&=\frac{ik_0}{4}\left[A(0,z)-A(\vec{x}',z)\right].
    \label{eqn:Markov}
\end{align}

In these differential coordinates the spectral coherency matrix then becomes
\begin{align}
    W_{ij}&(\vec{x},\vec{x}\,',z,\omega)
    \equiv \left\langle E_i\left(\vec{x}-\frac{\vec{x}\,'}{2},z,\omega\right)E_j^*\left((\vec{x}+\frac{\vec{x}\,'}{2},z,\omega\right)\right\rangle,~i,j=X,Y
    \label{eqn:specohere2}
\end{align}
Equation (\ref{eqn:parabolic5}) matches that of Charnotskii \cite{Charnotskii:16} (Equation 22 in the cited work) and states that the generalized Stokes parameters obey a parabolic equation of the same basic structure as the complex electric field amplitude in the deterministic, coherent case (see also, \cite{Fannjiang:05}). 

Now consider the spatial Fourier Transform of (\ref{eqn:specohere2}) with respect to the variable $\vec{x}\,'$ and the resulting Fourier Transform pair
\begin{align}
    \widehat{W}_{ij}(\vec{x},\vec{\xi},z,\omega)&=\int_{\R^2}W_{ij}(\vec{x},\vec{x}\,',z,\omega)e^{-i \vec{\xi}\cdot\vec{x}'}d\vec{x}\,'
    \nonumber \\
     W_{ij}(\vec{x},\vec{x}\,',z,\omega)&=\left(\frac{1}{2\pi}\right)^2\int_{\R^2}\widehat{W}_{ij}(\vec{x},\vec{\xi},z,\omega)e^{i \vec{\xi}\cdot\vec{x}\,'}d\vec{\xi}
     \label{eqn:FTpair}
\end{align}
where the hat $\widehat{W}$ denotes the spatial Fourier transform, the operator $\int_{\R^n}$ denotes the $n-$dimensional integral over the space of real numbers (e.g., $\int_{\R^2}\equiv \int_{-\infty}^{\infty}\int_{-\infty}^{\infty}$), and where the differential element in transverse real space and reciprocal 2-space are denoted by, respectively,
$d\vec{x}\,'\equiv\,dx'dy'$ and
$d\vec{\xi}\equiv\,d\xi_x d\xi_y$.
Note that, unlike in Eqn. (\ref{eqn:planewave}), we do not need to restrict the limits of integration in defining the Fourier Transform as the beam spatial correlations are assumed to be naturally limited in transverse extent, that is $|W_{ij}(\vec{x},\pm\infty,z,\omega)|\rightarrow 0$. The quantity $\widehat{W}_{ij}(\vec{x},\vec{\xi},z,\omega)$ is thus the Wigner transform  \cite{Wigner1932} associated with complex arguments $E_i(\vec{x},z,\omega),~E_j(\vec{x},z,\omega)$, $i,j\in X,Y$. 

In the expressions (\ref{eqn:FTpair}), 
$W_{ij}(\vec{x},\vec{x}',z,\omega)$ is the spatial covariance between the two field components $i,j$ as a function of separation $\vec{x}\,'$ in the transverse plane, at transverse position $\vec{x}$ and downrange position $z$, and within a small optical spectral range $\{\omega,\omega+d\omega\}$, while
$\widehat{W}_{ij}(\vec{x},\vec{\xi},z,\omega)$ is the covariance between Fourier amplitudes of electric field components $i,j$ associated with transverse spatial frequency $\vec{\xi}$, at downrange position $z$ and within a small optical spectral range $\{\omega,\omega+d\omega\}$.  Importantly, this latter quantity is proportional to the average phase difference between field components $i,~j$ with transverse separation $\vec{x}'$ and corresponding spatial frequency $\vec{\xi}$ (see Priestly section 9.1 \cite{Priestly:81}). 
 This point will be re-visited in section (\ref{sec:velocity}). 

We replace the spectral density matrix in coordinates $\vec{x},~\vec{x}\,'$ with its Wigner representation to obtain 
%\begin{align}
%    i2k_0\left(\frac{1}{2\pi}\right)\partial_z\int_{\R^2}\left(\begin{array}{c} \widehat{s_0'}\\\widehat{s_1'}\\\widehat{s_2'}\\\widehat{s_3'}\end{array}\right)e^{i\vec{\xi}\cdot\vec{x}'}d\vec{\xi}+2(\nabla_{X}\cdot\nabla_{X'})\left(\frac{1}{2\pi}\right)\int_{\R^2}\left(\begin{array}{c} \widehat{s_0'}\\\widehat{s_1'}\\\widehat{s_2'}\\\widehat{s_3'}\end{array}\right)e^{i\vec{\xi}\cdot\vec{x}'}d\vec{\xi}+2k_0^2\Phi(\vec{x},\vec{x}',z)\left(\begin{array}{c} s_0'\\s_1'\\s_2'\\s_3'\end{array}\right)&=\left(\begin{array}{c} 0\\0\\0\\0\end{array}\right)\nonumber \\
%    \frac{1}{2\pi}
%    \int_{\R^2}\left[\partial_z \left(\begin{array}{c} \widehat{s_0'}\\\widehat{s_1'}\\\widehat{s_2'}\\\widehat{s_3'}\end{array}\right)+\frac{1}{k_0}(\vec{\xi}\cdot\nabla_{X})\left(\begin{array}{c} \widehat{s_0'}\\\widehat{s_1'}\\\widehat{s_2'}\\\widehat{s_3'}\end{array}\right)-\int_{\R^2}ik_02\pi\Phi(\vec{x},\vec{x}',z)\left(\begin{array}{c} s_0'\\s_1'\\s_2'\\s_3'\end{array}\right)e^{-i\vec{\xi}\cdot \vec{x}'}d\vec{x}'\right]e^{i\vec{\xi}\cdot\vec{x}'}d\vec{\xi}&=\left(\begin{array}{c} 0\\0\\0\\0\end{array}\right)
%    \label{eqn:parabolic6}
%\end{align}
\begin{align}
   i2k_0\left(\frac{1}{2\pi}\right)^2\partial_z\int_{\R^2}\widehat{s}_\nu\,'e^{i\vec{\xi}\cdot\vec{x}'}d\vec{\xi}+2(\nabla_{X}\cdot\nabla_{X'})\left(\frac{1}{2\pi}\right)^2\int_{\R^2}\widehat{s}_\nu\,'e^{i\vec{\xi}\cdot\vec{x}'}d\vec{\xi}+2k_0^2\Phi(\vec{x},\vec{x}',z)s_\nu\,'&=0\nonumber \\
    \left(\frac{1}{2\pi}\right)^2
    \int_{\R^2}\left[\partial_z \widehat{s}_\nu\,'+\frac{1}{k_0}(\vec{\xi}\cdot\nabla_{X})\widehat{s}_\nu\,'-\int_{\R^2}ik_0\Phi(\vec{x},\vec{x}',z)s_\nu
    \,'e^{-i\vec{\xi}\cdot \vec{x}'}d\vec{x}'\right]e^{i\vec{\xi}\cdot\vec{x}'}d\vec{\xi}&=0,~\nu=0,\cdots,3
    \label{eqn:parabolic6}
    \end{align}
where now the generalized Stokes parameters $s_\nu\,'$ have been replaced by their respective spatial Fourier transforms $\widehat{s}_\nu\,'$ with respect to the differential coordinate $\vec{x}\,'$ and, hence, are written in terms of the Wigner distribution functions
\begin{align}
    \widehat{s}_0\,'(\vec{x},\vec{\xi},z,\omega)&=\widehat{W}_{XX}(\vec{x},\vec{\xi},z,\omega)+\widehat{W}_{YY}(\vec{x},\vec{\xi},z,\omega)\nonumber \\
\widehat{s}_1\,'(\vec{x},\vec{\xi},z,\omega)&=\widehat{W}_{XX}(\vec{x},\vec{\xi},z,\omega)-\widehat{W}_{YY}(\vec{x},\vec{\xi},z,\omega)\nonumber \\    \widehat{s}_2\,'(\vec{x},\vec{\xi},z,\omega)&=\widehat{W}_{XY}(\vec{x},\vec{\xi},z,\omega)+\widehat{W}_{YX}(\vec{x},\vec{\xi},z,\omega)\nonumber \\
 \widehat{s}_3\,'(\vec{x},\vec{\xi},z,\omega)&=i\left(\widehat{W}_{XY}(\vec{x},\vec{\xi},z,\omega)-\widehat{W}_{YX}(\vec{x},\vec{\xi},z,\omega)\right)
    \label{eqn:Luis}
\end{align}
which comprise the same result as Luis' \cite{Luis:07} Eq. (19). Recognizing that the term in square brackets must be zero in order to satisfy (\ref{eqn:parabolic6}), we obtain the vector transport equations,
\begin{align}
    \partial_z \widehat{\mathbfcal{S}}(\vec{x},\vec{\xi},z,\omega)+\frac{1}{k_0}(\vec{\xi}\cdot\nabla_{X})\widehat{\mathbfcal{S}}(\vec{x},\vec{\xi},z,\omega)&=\int_{\R^2}ik_0\Phi(\vec{x},\vec{x}',z)\mathbfcal{S}(\vec{x},\vec{x}',z,\omega)e^{-i\vec{\xi}\cdot \vec{x}'}d\vec{x}'\nonumber \\
    &=-\frac{ik_0}{(2\pi)^2}\int_{\R^4} \Phi(\vec{x},\vec{x}',z,\omega)\widehat{\mathbfcal{S}}(\vec{x},\vec{\xi'},z,\omega)e^{i(\vec{\xi}'-\vec{\xi})\cdot\vec{x}'}d\vec{\xi}'d\vec{x}'\nonumber \\
    &=-\frac{ik_0}{(2\pi)^2}\int_{\R^2} \Phi(\vec{x},\vec{\xi}-\vec{\xi}',z,\omega)\widehat{\mathbfcal{S}}(\vec{x},\vec{\xi'},z,\omega)d\vec{\xi}'
    \label{eqn:parabolic7}
\end{align}
where the 4-vector ${\bf \widehat{\mathbfcal{S}}}\equiv \left\{\widehat{s}_0\,',\widehat{s}_1\,',\widehat{s}_2\,',\widehat{s}_3\,'\right\}^T$ and $ \mathbfcal{S}$ is the associated inverse FT with respect to wavevector $\vec{\xi}$.  Equation (\ref{eqn:parabolic7}) thus denotes four separate transport equations, each governing one component $\widehat{s}\,'_\nu,~\nu=0\cdots 3$.  The term involving the vector potential can be written in several useful forms; the second line and matches the results of \cite{Yin:13} (Eqn. 2.4) and \cite{Jungel:06} (Eqn. 1.1), albeit with a different convention for the Fourier Transform, while the final convolution form is found in \cite{Cai:12} (Eqn. 2). 

Expression (\ref{eqn:parabolic7}) is the fundamental vector transport equation that governs the evolution of the generalized Stokes parameters in the direction of propagation, $z$.  In the scalar case (e.g., analysis of the 1D Schr\"{o}dinger equation \cite{Yin:13} or scalar optical wave propagation \cite{Friberg:86}, \cite{Petruccelli:13}), Eq. (\ref{eqn:parabolic7}) is often taken as the starting point in predicting the evolution of $\widehat{W}_{XX}(\vec{x},\vec{\xi},z,\omega)$.  Here we have extended these prior works to the full vector electric field in order to account for spatially dependent polarization.  Related works on the full vector transport equations for the Wigner distribution include \cite{Ryzhik:96} (Eq. 1.6), and \cite{Manfredi:21}.  However we believe this work to be the first extension of such analyses to partially polarized optical fields.

\section{Partially-Coherent, Vector Transport of Intensity}
In this section we extend the transport-of-intensity equations (TIEs) from the partially coherent scalar case, as described in \cite{Petruccelli:13,Zuo:15}, to the partially-coherent vector case.   To arrive at such a model, we follow an approach used in quantum mechanics for obtaining the so-called ``hydrodynamic'' model, so named for the resemblance of the model to the continuity and momentum equations commonly found in fluid mechanics. 
Specifically, we take ``moments'' of Eq. (\ref{eqn:parabolic7}) with respect to the wavevector.  The first two such moments are obtained by multiplying (\ref{eqn:parabolic7}) by unity and $\vec{\xi}$ respectively, and then integrating over the wavevector.  The resulting expressions govern the conservation of mass (intensity), momentum (phase), and energy of the system.  This approach has been used in analysis of the 1-D Schr\"{o}dinger equation (see \cite{Degond:03,Jungel:06,Cai:12}) and has only very recently been extended to the 2-D Pauli equation \cite{Manfredi:21}.  While the dominant application area has been quantum mechanics, Marcuvitz \cite{Marcuvitz:80} has suggested the approach as a general technique for studying wave propagation in the scalar case.
 
In this work, we extend this ``method of moments'' to the vector transport equation (\ref{eqn:parabolic7}) for partially coherent propagation problems in optics.  The first such conservation equation is obtained by directly integrating (\ref{eqn:parabolic7}) (that is, taking the ``zeroth'' moment) which yields 
\begin{align}
\partial_z\int_{\R^2} \widehat{\mathbfcal{S}}(\vec{x},\vec{\xi},z,\omega) 
d{\vec{\xi}}+\nabla_X\cdot \frac{1}{k_0}\int_{\R^2}\vec{\xi}~\widehat{\mathbfcal{S}}(\vec{x},\vec{\xi},z,\omega)d\vec{\xi}={\bf 0}.
\label{eqn:rawtransport}
\end{align}
To see that the Wigner potential term integrates to zero, replace the generalized Stokes parameters by their inverse spatial Fourier transforms (in frequency variable $\vec{\xi}'$), in which case integrating the right hand side of (\ref{eqn:parabolic7}) over wavevector $\vec{\xi}$ can be carried out as
\begin{align}
    &=\int_{\R^6} \frac{ik_0}{(2\pi)^2} \widehat{\mathbfcal{S}}(\vec{x},\vec{\xi}\,',z,\omega)e^{i\vec{\xi}\,'\cdot \vec{x}'}\Phi(\vec{x},\vec{x}\,',z)e^{-i\vec{\xi}\cdot\vec{x}\,'}d\vec{x}\,'d\vec{\xi}\,'d\vec{\xi}\nonumber \\
    &=\frac{ik_0}{(2\pi)^2}\int_{\R^4}\widehat{\mathbfcal{S}}(\vec{x},\vec{\xi}\,',z,\omega)e^{i\vec{\xi}\,'\cdot\vec{x}\,'}d\vec{\xi}\,'\int_{\R^2}e^{-i\vec{\xi}\cdot\vec{x}\,'}d\vec{\xi}\Phi(\vec{x},\vec{x}\,')d\vec{x}\,'\nonumber \\
    &=\frac{ik_0}{(2\pi)^2}\int_{R^4}\widehat{\mathbfcal{S}}(\vec{x},\vec{\xi}\,',z,\omega)e^{i\vec{\xi}\,'\cdot\vec{x}\,'}d\vec{\xi}\,'\delta(\vec{x}\,')\Phi(\vec{x},\vec{x}\,')d\vec{x}\,'\nonumber \\
    &=\frac{ik_0}{(2\pi)^2}\Phi(\vec{x},0)\int_{\R^2}\widehat{\mathbfcal{S}}(\vec{x},\vec{\xi}\,',z,\omega)d\vec{\xi}\,'=0.
\end{align}
In the last step we have leveraged the definition of $\Phi(\vec{x},\vec{x}\,',z)$ (\ref{eqn:vecPot},\ref{eqn:Markov}) which holds that when $\vec{x}'=0$, the potential vanishes, that is, $\Phi(\vec{x},0)=0$.  This statement is true whether we treat the dielectric constant fluctuations as a deterministic or a random process under the Markov approximation (\ref{eqn:Markov}).  Finally, to put (\ref{eqn:rawtransport}) in the same form as the coherent Transport of Intensity Equation (TIE) \cite{Nichols:22}, we require the definitions
\begin{align}
\boldsymbol{\rho}(\vec{x},z,\omega)&\equiv\int_{\R^2}\widehat{\mathbfcal{S}}(\vec{x},\vec{\xi},z,\omega)d\vec{\xi}\nonumber \\
    \frac{1}{k_0}\int_{\R^2} \vec{\xi}\widehat{\mathbfcal{S}}(\vec{x},\vec{\xi},z,\omega)d\vec{\xi}&\equiv\boldsymbol{\rho}(\vec{x},z,\omega)\circ\vec{\bf v}(\vec{x},z,\omega)
    \label{eqn:velnum}
\end{align}
\\[10pt]
where $\circ$ denotes the Hadamard product. Equation (\ref{eqn:velnum}) therefore defines four generalized intensities $\rho_\nu(\vec{x},z,\omega)\equiv \int_{\R^2} \widehat{s}_\nu\,'(\vec{x},\vec{\xi},z,\omega)d\vec{\xi},~\nu=0,1,2,3$.
In addition, (\ref{eqn:velnum}) implicitly defines each element of $\vec{\bf{ v}}(\vec{x},z,\omega)$ as
\begin{align}
\vec{v}_\nu(\vec{x},z,\omega)
\equiv\frac{\int_{\R^2}\vec{\xi}~\widehat{s}_\nu\,'(\vec{x},\vec{\xi},z,\omega)d\vec{\xi}}{k_0 \rho_\nu(\vec{x},z,\omega)}=\frac{\int_{\R^2}\vec{\xi}~\widehat{s}_\nu\,'(\vec{x},\vec{\xi},z,\omega)d\vec{\xi}}{k_0 \int_{\R^2}\widehat{s}_\nu\,'(\vec{x},\vec{\xi},z,\omega)d\vec{\xi}}~,~\nu=0,1,2,3
\label{eqn:avgvel}
\end{align}
in which case (\ref{eqn:rawtransport}) becomes
\begin{align}
    \partial_z\boldsymbol{\rho}(\vec{x},z,\omega)+\nabla_X\cdot \left[\boldsymbol{\rho}(\vec{x},z,\omega)\circ\vec{\bf v}(\vec{x},z,\omega)\right]=0,
    \label{eqn:vec_cont}
\end{align}
which is the vector TIE for partially coherent light. Note that whereas in our prior analysis \cite{Nichols:22}, the above expression was a single scalar equation, in this case both $\boldsymbol{\rho}(\vec{x},z,\omega)$ and the product $\boldsymbol{\rho}(\vec{x},z,\omega) \circ\vec{\bf v}(\vec{x},z,\omega)$ are column vectors each comprising four elements. Thus $\rho_\nu$ and $v_\nu$ represent, respectively, the $\nu-$th component of the Stokes 4-vector $(\nu=0,1,2,3)$ and, as we will show in section (\ref{sec:velocity}), the rate of change with $z$ of the transverse location of the $\nu-$th component of the Stokes 4-vector. Hence, all eight components $\rho_\nu$ and $v_\nu$ are real, the total beam intensity $\rho_0$ must be non-negative, and the remaining seven components can be positive, negative, or zero.  Just as for Stokes parameters for monochromatic light, the components of the Stokes 3-vector $(\rho_1, \rho_2, \rho_3)$ are proportional to intensity but are not limited to non-negative values.

\subsection{Mixture Model for Partially Coherent Stokes Parameters}

Equations (\ref{eqn:vec_cont}) are transport equations governing each of the generalized Stokes parameters during propagation in a (possibly) inhomogeneous medium.  By definition, the Stokes parameters were defined as an expectation over realizations of the $X,~Y$ components of the electric field Fourier amplitudes, $E_X(\vec{x},z,\omega),~E_Y(\vec{x},z,\omega)$. 
 Both the quantity being transported, $\boldsymbol{\rho}(\vec{x},z,\omega)$ and the ``velocity'' of that transport, $  \vec{\bf v}(\vec{x},z,\omega)$ carry simple physical interpretations when placed in the context of a particular model for the electric field as will be shown next.

 To aid in this interpretation, recall the electric field ``transverse wave'' model (\ref{eqn:planewave}).  Until now we have imposed no further model structure on the form of  $E_X(\vec{x},z,\omega),~E_Y(\vec{x},z,\omega)$.  Leveraging prior  work \cite{Nichols:22} for linearly-polarized, coherent beams, let the complex Fourier amplitudes take the form 
\begin{align}
E_X(\vec{x},z,\omega)&=\rho_m^{1/2}(\vec{x},z,\omega)e^{i\phi_{m}(\vec{x},z,\omega)}\cos(\gamma_m(\vec{x},z,\omega))\nonumber \\
E_Y(\vec{x},z,\omega)&=\rho^{1/2}_{m}
(\vec{x},z,\omega)e^{i\phi_{m}(\vec{x},z,\omega)}\sin(\gamma_m(\vec{x},z,\omega))
\label{eqn:modelstruct}
\end{align}
where the subscript $m$ denotes ``model''.  The electric field is therefore described by a magnitude, phase, and {\it spatially dependent} linear polarization angle $\gamma_m(\vec{x},z,\omega)$ which governs the ratio of Fourier amplitudes in the $X$ and $Y$ directions.  It must be noted, however, that the model structure (\ref{eqn:modelstruct}) {\it presumes} a polarized beam and should therefore be viewed as {\it conditional} on the beam being completely polarized. Such a model is clearly not appropriate for unpolarized light. 

While the first Stokes parameter $\widehat{s}_0$ is independent of the state of polarization, the same cannot be said of $\widehat{s}_\nu,~\nu=1\cdots 3$. For partially polarized light it is therefore appropriate to use the mixture model \cite{Schaefer:07}
\begin{align}
\left(
\begin{array}{c} s_0'(\vec{x},\vec{x}',z,\omega)\\s_1'(\vec{x},\vec{x}',z,\omega)\\s_2'(\vec{x},\vec{x}',z,\omega)\\s_3'(\vec{x},\vec{x}',z,\omega) \end{array}
\right)
&=
\left[1-\mathcal{P}\right]\left(\begin{array}{c} s_0'(\vec{x},\vec{x}',z,\omega)\\0\\0\\0 \end{array}
\right)+\mathcal{P}\left(\begin{array}{c} s_0'(\vec{x},\vec{x}',z,\omega)\\s_1'(\vec{x},\vec{x}',z,\omega|\mathcal{P})\\s_2'(\vec{x},\vec{x}',z,\omega|\mathcal{P})\\s_3'(\vec{x},\vec{x}',z,\omega|\mathcal{P})\end{array}\right)
\label{eqn:breakdown}
\end{align}
where the weighting is given by the Degree Of Polarization (DOP) 
\begin{align}
\mathcal{P}(\vec{x},z,\omega)&=\dfrac
{\sqrt{(s_1)^2+(s_2)^2+(s_3)^2}}
{s_0}\Big|_{\vec{x}_1=\vec{x}_2=\vec{x}}=~\sqrt{1-\frac{4\mathrm{Det}\left[{\bf W}\right]}{\mathrm{Tr}\left[{\bf W}\right]^2}}\Big|_{\vec{x}_1=\vec{x}_2=\vec{x}}.
\label{eqn:DOP}
\end{align}
Equation (\ref{eqn:DOP}) thus defines a positive quantity $0\le \mathcal{P}(\vec{x},z,\omega)\le 1$ which specifies the fraction of the beam intensity at location $\{\vec{x},z\}$ and within an infinitesimal range of optical frequencies $\{\omega,\omega+d\omega\}$ that is fully polarized.  %Both forms of (\ref{eqn:DOP}) can be verified as equivalent by considering Eqn. (\ref{eqn:specohereAlt}). 
The degree to which a beam is polarized is therefore directly related to coherence among the components of the optical field at a given transverse location. We note also that while $\mathcal{P}(\vec{x},z,\omega)$ varies spatially and with frequency, as defined it does not depend on the separation $\vec{x}\,'$ between two points on the beam face.  

Given the above definitions, we can better understand the quantities in the partially coherent TIE (\ref{eqn:vec_cont}) as will be described next in sections (\ref{sec:pcTIE}) and (\ref{sec:velocity}).  These definitions will also allow us to define a polarization gradient in the partially coherent case as will be shown in section (\ref{sec:pcAng}).

\subsection{Interpretation of the Partially Coherent Vector TIE \label{sec:pcTIE}}

To understand the meaning of $\boldsymbol{\rho}(\vec{x},z,\omega)$ we expand the first component of $\widehat{\mathbfcal{S}}(\vec{x},\vec{\xi},z,\omega)$ which is just $\widehat{s}_0\,'(\vec{x},\vec{\xi},z,\omega)$ and use the fact that the Wigner distributions can be expressed as Fourier transforms to obtain
\begin{align}
    \rho_0(\vec{x},z,\omega)&=\int_{R^2} \widehat{s}_0\,'(\vec{x},\vec{\xi},z,\omega)d\vec{\xi}=\int_{R^4} \left(W_{XX}(\vec{x},\vec{x}',z,\omega)+W_{YY}(\vec{x},\vec{x}',z,\omega)\right)e^{-i\vec{\xi}\cdot\vec{x}'}d\vec{x}\,'d\vec{\xi}\nonumber \\
    &=\int_{\R^2} \left(W_{XX}(\vec{x},\vec{x}\,',z,\omega)+W_{YY}(\vec{x},\vec{x}\,',z,\omega)\right)\delta(\vec{x}\,')d\vec{x}\,'\nonumber \\
    &=S_{XX}(\vec{x},z,\omega)+S_{YY}(\vec{x},z,\omega)
    \label{eqn:rcomp1}
\end{align}
where we have invoked the definition of the delta function as the Fourier Transform of unity.  Recall also that by definition, in transformed coordinates, at $\vec{x}\,'=0$,
\begin{align}
    W_{ij}(\vec{x},0,z,\omega)
    &=S_{ij}(\vec{x},z,\omega),~i,j=X,Y
    \label{eqn:specohere3}
\end{align}
so that the first component of $\boldsymbol{\rho}(\vec{x},z,\omega)$ is by definition the auto-spectral density of the field at transverse location $\vec{x},~z$ as demonstrated by  (\ref{eqn:rcomp1}).
A similar analysis of the other components of $\boldsymbol{\rho}(\vec{x},z,\omega)$ shows that the full four-component Stokes vector is given by
\begin{align}
    \boldsymbol{\rho}(\vec{x},z,\omega)&=\int_{\R^2} \left(\begin{array}{c} \widehat{s_0'}\\ \mathcal{P}\widehat{s_1'}\\ \mathcal{P}\widehat{s_2'}\\ \mathcal{P}\widehat{s_3'}\end{array}\right)d\vec{\xi}=\left(\begin{array}{c}
    S_{XX}(\vec{x},z,\omega)+S_{YY}(\vec{x},z,\omega)\\
    \mathcal{P}\left(S_{XX}(\vec{x},z,\omega)-S_{YY}(\vec{x},z,\omega)\right)\\
    \mathcal{P}\left(S_{XY}(\vec{x},z,\omega)+S_{YX}(\vec{x},z,\omega)\right)\\
    i\mathcal{P}\left(S_{XY}(\vec{x},z,\omega)-S_{YX}(\vec{x},z,\omega)\right) \end{array}
    \right)=\left(\begin{array}{c} s_0\\ \mathcal{P}s_1\\ \mathcal{P}s_2\\ \mathcal{P}s_3\end{array}\right)_{\vec{x}_1=\vec{x}_2\equiv \vec{x}}
    \label{eqn:rhocomps}
\end{align}
Thus, the quantities being transported can be written simply as sums and differences of the transverse spectral densities that is, the generalized Stokes parameters for partially coherent light.

For example, substituting the model (\ref{eqn:modelstruct}) for the first Stokes parameter
\begin{align}
s_0&=\left\langle E_X(\vec{x},z,\omega)E_X^*(\vec{x},z,\omega)\right\rangle+\left\langle E_Y(\vec{x},z,\omega)E_Y^*(\vec{x},z,\omega)\right\rangle\nonumber \\
&=\left\langle \rho_m(\vec{x},z,\omega)\cos^2(\gamma_m(\vec{x},z,\omega)+\rho_m(\vec{x},z,\omega)\sin^2(\gamma_m(\vec{x},z,\omega)\right\rangle \nonumber \\
&=\left\langle\rho_m(\vec{x},z,\omega)\right\rangle \equiv \rho_0(\vec{x},z,\omega).
\end{align}
Hence the first generalized Stokes parameter $\rho_0(\vec{x},z,\omega)$ is proportional to the expected beam spectral intensity at location $\{\vec{x},z\}$ in our model. Performing the same substitution for the remaining components in (\ref{eqn:rhocomps}) we have 
\begin{align}
    \boldsymbol{\rho}(\vec{x},z,\omega)&=\left(\begin{array}{c}
\left\langle\rho_m(\vec{x},z,\omega)\right\rangle \\
\mathcal{P}(\vec{x},z,\omega)\left\langle\rho_m(\vec{x},z,\omega)\cos(2\gamma_m(\vec{x},z,\omega))\right\rangle \\    \mathcal{P}(\vec{x},z,\omega)\left\langle\rho_m(\vec{x},z,\omega)\sin(2\gamma_m(\vec{x},z,\omega))\right\rangle \\
    0\end{array}
    \right).
\end{align}
The model structure (\ref{eqn:modelstruct}) suggests a simple and well-known interpretation of the generalized Stokes parameters for linearly polarized light.  However, for unpolarized light only the first component of $\boldsymbol{\rho}(\vec{x},z,\omega)$ is non-zero.  
The final intensity component is given by $\rho_3(\vec{x},z,\omega)=0$ due to the fact that we are considering only linear polarization.

%\textcolor{red}{the time-average total energy spectral(?) density density in this TEM wave is}
%\begin{equation}\nonumber
%\color{red} u(\vec{x},z,\omega)=\epsilon_0\rho_0(\vec{x},z,\omega)
%\end{equation}
%\textcolor{red}{(There might be  a bandwidth factor missing here.)}

\subsection{Interpretation of the Transport Velocity \label{sec:velocity}}

The components of $\vec{\bf v}(\vec{x},z,\omega)$ in Eq. (\ref{eqn:vec_cont}) are referred to as ``velocities'' as they were noted in the coherent case to govern the change in optical path in the transverse direction per unit change in the direction of propagation \cite{Nichols:19,Nichols:22}. We present a basic derivation supporting this interpretation in Appendix (\ref{app:velocity}).  This terminology is also in keeping with the wave mechanics interpretation of the transport equation, where the ``transport'' of intensity occurs in the transverse plane as the beam progresses in $z$.  The first velocity term in this partially coherent case is obtained via (\ref{eqn:avgvel}),
\begin{align}    \vec{v}_0(\vec{x},z,\omega)&=\frac{1}{k_0}\frac{\int_{\R^2}\vec{\xi}~\widehat{s'_0}(\vec{x},\vec{\xi},z,\omega)d\vec{\xi}}{ \int_{\R^2}\widehat{s'_0}(\vec{x},\vec{\xi},z,\omega)d\vec{\xi}}=\frac{\int_{\R^2}\vec{\xi}\left[ \widehat{W}_{XX}(\vec{x},\vec{\xi},z,\omega)+\widehat{W}_{YY}(\vec{x},\vec{\xi},z,\omega)\right]d\vec{\xi}}{k_0\rho_0(\vec{x},z,\omega)}
    \label{eqn:WignerPhase}
\end{align}
The non-negativity of $\widehat{s'_0}$ allows us to interpret (\ref{eqn:WignerPhase}) as an average spatial frequency in the transverse plane (see e.g., Boashash \cite{Boashash92P1}) normalized by $k_0$.  In fact, in \cite{Boashash92P1} it was noted that the generalized definition of frequency is an average change in phase per unit change in the independent variable, which in the present context corresponds to the transverse distance $\vec{x}$.  Thus, (\ref{eqn:WignerPhase}) implicitly defines a phase gradient 
\begin{align}
\nabla_X\phi(\vec{x},z,\omega)\equiv 
\frac{\int_{\R^2}\vec{\xi}~\widehat{s'_0}(\vec{x},\vec{\xi},z,\omega)d\vec{\xi}}{ \int_{\R^2}\widehat{s'_0}(\vec{x},\vec{\xi},z,\omega)d\vec{\xi}}
\label{eqn:WignerPhase2}
\end{align}
so that $\vec{v}_0(\vec{x},z,\omega)=k_0^{-1}\nabla_X\phi(\vec{x},z,\omega)$, which is precisely the definition for velocity found in the monochromatic case \cite{Nichols:22}.  This definition of generalized phase was also used in \cite{Zhou:15} for scalar, partially coherent electric fields.  Here, we have shown this definition to be appropriate for  vector electric fields as well.  Indeed, we will show momentarily that these definitions are not only appropriate for transport of intensity, $\rho_0(\vec{x},z,\omega)$, but for transport of two other generalized Stokes parameters.  

Now consider again the mixture model (\ref{eqn:breakdown}).  Substituting these definitions into (\ref{eqn:WignerPhase}) yields for the denominator $S_{XX}(\vec{x},z,\omega)+S_{YY}(\vec{x},z,\omega)=\rho_0(\vec{x},z,\omega)$.  For the numerator, considering first the polarized portion, 
we can expand using Eqn. (\ref{eqn:specohere2}) to give
\begin{align}
    \int_{\R^2} &\vec{\xi}\left\{\widehat{W}_{XX}(\vec{x},\vec{\xi},z,\omega)+\widehat{W}_{YY}(\vec{x},\vec{\xi},z,\omega)\right\}d\vec{\xi}\nonumber \\
    &=\int_{\R^4} \vec{\xi}\left\{\left\langle E_X(\vec{x}-\vec{x'}/2,z,\omega)E_X^{*}(\vec{x}+\vec{x'}/2,z,\omega)\right\rangle+\left\langle E_Y(\vec{x}-\vec{x'}/2,z,\omega)E_Y^{*}(\vec{x}+\vec{x'}/2,z,\omega)\right\rangle\right\}e^{-i\vec{\xi}\cdot\vec{x}'}d\vec{\xi}d\vec{x}'\nonumber \\
    &=\int_{\R^2} \left\{\left
    \langle E_X(\vec{x}-\vec{x'}/2,z,\omega)E_X^{*}(\vec{x}+\vec{x'}/2,z,\omega)\right\rangle+\left\langle E_Y(\vec{x}-\vec{x'}/2,z,\omega)E_Y^{*}(\vec{x}+\vec{x'}/2,z,\omega)\right\rangle\right\}\int_{\R^2}\vec{\xi}e^{-i\vec{\xi}\cdot\vec{x}'}d\vec{\xi}d\vec{x}'\nonumber \\
    &=i\int_{\R^2} \left\{\left\langle E_X(\vec{x}-\vec{x'}/2,z,\omega)E_X^{*}(\vec{x}+\vec{x'}/2,z,\omega)\right\rangle+\left\langle E_Y(\vec{x}-\vec{x'}/2,z,\omega)E_Y^{*}(\vec{x}+\vec{x'}/2,z,\omega)\right\rangle\right\}\delta'(\vec{x}')d\vec{x}'\nonumber \\
    &=i\int_{\R^2} \nabla_{X'}\left\{\left\langle E_X(\vec{x}-\vec{x'}/2,z,\omega)E_X^{*}(\vec{x}+\vec{x'}/2,z,\omega)\right\rangle+\left\langle E_Y(\vec{x}-\vec{x'}/2,z,\omega)E_Y^{*}(\vec{x}+\vec{x'}/2,z,\omega)\right\rangle\right\}\delta(\vec{x}')d\vec{x}'\nonumber \\
    &=i\nabla_{X'}\left\{\left\langle E_X(\vec{x}-\vec{x'}/2,z,\omega)E_X^{*}(\vec{x}+\vec{x'}/2,z,\omega)\right\rangle+\left\langle E_Y(\vec{x}-\vec{x'}/2,z,\omega)E_Y^{*}(\vec{x}+\vec{x'}/2,z,\omega)\right\rangle\right\}\Bigg|_{\vec{x}'\rightarrow 0}\nonumber \\
    &=\left\langle\rho_{m}(\vec{x},z)\nabla_X\phi_{m}(\vec{x},z)\right\rangle.
    \label{eqn:veldef}
\end{align}
where the final line substitutes Eq. (\ref{eqn:modelstruct}) for $E_X,~E_Y$ and takes the required gradient $\nabla_{X'}$.  In the second to last line we leveraged the identity in Eq. (7) of \cite{Boykin:03} concerning integrals over derivatives of the delta function. The end result is that under the mixture model, and assuming amplitude and transverse phase gradient are statistically independent,
\begin{align}
    \left\langle\nabla_X\phi_{m}(\vec{x},z,\omega)\right\rangle&=\nabla_X\phi(\vec{x},z,\omega)\nonumber \\
    \frac{1}{k_0}\left\langle\nabla_X\phi_{m}(\vec{x},z,\omega)\right\rangle&=\vec{v}_0(\vec{x},z,\omega)
\end{align}
The definition (\ref{eqn:WignerPhase2}) therefore can be interpreted as the {\it expected} model phase and the corresponding ``velocity'' as the expected change in optical path in the transverse direction per unit change in the direction of propagation.  Notably, this result is independent of the DOP. 
 We may perform a similar analysis on the remaining components of $\vec{\bf v}(\vec{x},z,\omega)$, finding that
\begin{align}
\vec{v}_1(\vec{x},z,\omega)&=\frac{\int_{\R^2} \vec{\xi} ~\widehat{s_1'}(\vec{x},\vec{\xi},z,\omega)d\vec{\xi}}{k_0{\rho}_1(\vec{x},z,\omega)}= \frac{\mathcal{P}(\vec{x},z,\omega)\left\langle\rho_{m}\nabla_X\phi_{m}(\vec{x},z,\omega)\cos(2\gamma_m(\vec{x},z,\omega))\right\rangle}{k_0\mathcal{P}(\vec{x},z,\omega)\left\langle\rho_{m}\cos(2\gamma_m(\vec{x},z,\omega))\right\rangle}=\vec{v}_0(\vec{x},z,\omega)\nonumber \\
\vec{v}_2(\vec{x},z,\omega)&=\frac{\int_{\R^2} \vec{\xi} ~\widehat{s_2'}(\vec{x},\vec{\xi},z,\omega)d\vec{\xi}}{k_0{\rho}_2(\vec{x},z,\omega)}= \frac{\mathcal{P}(\vec{x},z,\omega)\left\langle\rho_{m}\nabla_X\phi_{m}(\vec{x},z,\omega)\sin(2\gamma_m(\vec{x},z,\omega))\right\rangle}{k_0\mathcal{P}(\vec{x},z,\omega)\left\langle\rho_{m}\sin(2\gamma_m(\vec{x},z,\omega))\right\rangle}=\vec{v}_0(\vec{x},z,\omega).
\end{align}
Thus, in the case of linear polarization, all three non-zero generalized Stokes parameters are transported with the same transverse velocity.
\begin{align}
    \partial_z\rho_\nu(\vec{x},z,\omega)&+\nabla_X\cdot \left[\rho_\nu(\vec{x},z,\omega)v_0(\vec{x},z,\omega)\right]=0,~\nu=0,1,2.
    \label{eqn:transport2}
\end{align}
This is sensible as it suggests that, in expectation, both the beam intensity and properties related to the state of linear polarization are moving together in the transverse plane during propagation.  Importantly, this also suggests that while the transport model requires an equation governing $\vec{v}_0$ (see section \ref{sec:momentum}), it does not require separate equations for $\vec{v}_1,~\vec{v}_2$.    

It it worth mentioning the relationship between the velocity vector and the well-known Poynting vector. Begin by noting that the numerator in (\ref{eqn:avgvel}) is of the form of the average Poynting vector for partially coherent light,
\begin{align}
P(\vec{x},z)\equiv \frac{1}{k_0}\int_{\R^2}\vec{\xi}~\widehat{s'_0}(\vec{x},\vec{\xi},z,\omega)d\vec{\xi},
\end{align}
so that $\tilde{P}(\vec{x},z)\equiv \vec{v}(\vec{x},z,\omega)$ is, in fact, the normalized Poynting vector as defined in \cite{Paganin:98} (Eq. 8 of the cited work).  Also in \cite{Paganin:98} is was suggested that the normalized Poynting vector be decomposed via the Helmholtz decomposition theorem as
\begin{align}
\vec{v}(\vec{x},z,\omega)&\equiv \vec{v}_S(\vec{x},z,\omega)+\vec{v}_V(\vec{x},z,\omega)\nonumber \\
&=k_0^{-1}\nabla_X\phi_S(\vec{x},z,\omega)+k_0^{-1}\nabla_X\times \vec{\phi}_V(\vec{x},z,\omega)
\end{align}
where $S$ and $V$ refer to scalar and vector contributions, respectively. If we let the scalar phase be $\phi_S(\vec{x},z,\omega)=\phi_m(\vec{x},z,\omega)$ and define 
\begin{align}
  \phi_V(\vec{x},z,\omega)=\left\{\frac{d\gamma(\vec{x},\omega)}{dy} z,-\frac{d\gamma(\vec{x},\omega)}{dx}z,0\right\}  
\end{align} 
then we have that $\vec{\Omega}(\vec{x},z,\omega)=[\nabla\times \phi_V(\vec{x},z,\omega)]_X$. 
The partially coherent vector TIE would then be written
\begin{align}
    \partial_z\rho(\vec{x},z,\omega)+\nabla_X\cdot\left[\rho(\vec{x},z,\omega)\vec{v}_S(\vec{x},z,\omega)\right]+\nabla_X\cdot\left[\rho(\vec{x},z,\omega)\vec{v}_V(\vec{x},z,\omega)\right]=0
\end{align}
which would clearly suggest that $\nabla_X\cdot \left[\rho(\vec{x},z,\omega)\vec{\Omega}(\vec{x},z,\omega)\right]=0$ given Eq. (\ref{eqn:vec_cont}).  In fact, we will show in the next section that this relationship indeed holds for linearly polarized light.

\subsection{Definition and Interpretation of the Polarization Angle Gradient \label{sec:pcAng}}

Since we are considering a linearly polarized vector beam $\rho_3(\vec{x},z,\omega)=0$.  However, although the first term in (\ref{eqn:rawtransport}) is zero, the second term inside the transverse divergence operator is (following the same model interpretation and simplification as in  (\ref{eqn:veldef}))
\begin{align}
\frac{1}{k_0}\int_{\R^2} \vec{\xi} ~\widehat{s_3'}(\vec{x},\vec{\xi},z,\omega)d\vec{\xi}&= -\left\langle\rho_{m}(\vec{x},z,\omega)\frac{\mathcal{P}(\vec{x},z,\omega)}{k_0}\nabla_X\gamma_m(\vec{x},z,\omega)\right\rangle=-\rho_0(\vec{x},z,\omega)\vec{\Omega}(\vec{x},z,\omega)
\label{eqn:s3moment}
\end{align}
so that by (\ref{eqn:rawtransport}) we have the condition
\begin{align}
\nabla_X\cdot\left[\rho_0(\vec{x},z,\omega)\vec{\Omega}(\vec{x},z,\omega)\right]=0
\label{eqn:divgrad}
\end{align}
as we implied at the end of the previous section.  This relationship was also found to hold in the coherent case \cite{Nichols:22} and will be leveraged later in the derivation.
However, also note that in arriving at (\ref{eqn:divgrad}) we have produced a definition for the spatial gradient of the polarization angle that is appropriate for partially coherent light, given by 
\begin{align}
\vec{\Omega}(\vec{x},z,\omega)&=
\frac{-i\mathcal{P}(\vec{x},z,\omega)\int_{\R^2}\vec{\xi} \left[\widehat{W}_{XY}(\vec{x},\vec{\xi},z,\omega)-\widehat{W}_{YX}(\vec{x},\vec{\xi},z,\omega)\right]d\vec{\xi}}{k_0\rho_0(\vec{x},z,\omega)}=\frac{\mathcal{P}(\vec{x},z,\omega)}{k_0}\left\langle\nabla_X\gamma_m(\vec{x},z,\omega)\right\rangle.
\label{eqn:gradangledef2}
\end{align}
The DOP is appropriately part of the definition (\ref{eqn:gradangledef2}) as one cannot define $\vec{\Omega}$ for an unpolarized beam.  Note also that in accordance with the works of Salem {\it et al.} \cite{Salem:04} and Korotkova {\it et al.} \cite{Korotkova:04} we allow that the DOP can vary in frequency $\omega$, as well as the transverse location $\vec{x}$, and the direction of propagation $z$.  

Just as with the normalized phase gradient, Eqn. (\ref{eqn:gradangledef2}) is expressed as a suitable average wavevector over the distribution $\widehat{s_3'}$.  Similarly, it can be interpreted as a transverse, spatial gradient in the angle (phase) between the $X$ and $Y$ components.  In the limit of a fully coherent beam we recover our deterministic definition.  Importantly, as the beam de-polarizes we will see the terms involving $\vec{\Omega}(\vec{x},z,\omega)$ vanish, a point we elaborate on in the next section.

We note that a definition of polarization angle for partially coherent beams was proposed previously by Korotkova {\it et al.} \cite{Wolf:05},~\cite{Korotkova:08}
\begin{align}
    \theta(\vec{x},z,\omega)&\equiv
    \frac{1}{2}\arctan\left(\frac{2\textrm{Re}\left\{S_{XY}(\vec{x},z,\omega)\right\}}{S_{XX}(\vec{x},z,\omega)-S_{YY}(\vec{x},z,\omega)}\right).
    \label{eqn:angle}
\end{align}
This definition is somewhat consistent with our model structure (\ref{eqn:modelstruct}) in the sense that substituting the polarized electric field (\ref{eqn:modelstruct}) into (\ref{eqn:angle}) gives
\begin{align}
\theta(\vec{x},z,\omega)&\equiv\frac{1}{2}\arctan\left(\frac{s_2(\vec{x},z,\omega)}{s_1(\vec{x},z,\omega)}\right)=\frac{1}{2}\arctan\left(\frac{\langle \sin(2\gamma_m(\vec{x},z,\omega))\rangle}{\langle \cos(2\gamma_m(\vec{x},z,\omega))\rangle}\right).
\label{eqn:angle2}
\end{align}
However, this definition only yields our model polarization angle in the deterministic case as the ratio of expectations of the $\sin,\cos$ terms does not equal the mean of the ratio in general.  Moreover, the DOP does not appear explicitly in the expression as it does in (\ref{eqn:gradangledef2}) and instead must be viewed as a part of the quantity $S_{XY}(\vec{x},z,\omega)$.

More importantly, as we will see in section (\ref{sec:momentum}), the quantity that governs the transverse movement of intensity is the polarization angle {\it gradient} as opposed to the angle itself.  As we have just shown, the gradient is obtained via the first moment, Eq. (\ref{eqn:s3moment}). The resulting expression (\ref{eqn:gradangledef2}) accounts for the spatial frequencies $\vec{\xi}$ and yields directly the expected value of polarization angle gradient. On the other hand, differentiating (\ref{eqn:angle}) with respect to the transverse coordinates yields
\begin{align}
   \nabla_X\theta&(\vec{x},z,\omega)\nonumber \\
    &=\frac{[S_{XX}(\vec{x},z,\omega)-S_{YY}(\vec{x},z,\omega)]\nabla_X S_{XY}(\vec{x},z,\omega)-S_{XY}(\vec{x},z,\omega)\nabla_X[S_{XX}(\vec{x},z,\omega)-S_{YY}(\vec{x},z,\omega)]}{4S_{XY}(\vec{x},z,\omega)^2+[S_{XX}(\vec{x},z,\omega)-S_{YY}(\vec{x},z,\omega)]^2}.
    \label{eqn:anglegrad}
\end{align}
As with Eq. (\ref{eqn:angle2}), this definition of phase gradient results in products and sums of expectations of $\sin,~\cos$ functions so that the result equals $\langle \nabla_X\gamma_m(\vec{x},z,\omega)\rangle$ only in the deterministic case where no averaging operations are performed.

Finally, we note that our definition (\ref{eqn:gradangledef2}) arose quite naturally as a consequence of 1) the paraxial transport equation for the generalized Stokes parameters, Eq. (\ref{eqn:rawtransport}) and 2) the model Eq. (\ref{eqn:modelstruct}) governing the expected amplitude, phase, and polarization angle of the electric field at location $(\vec{x},z)$ and frequency $\omega$.  The definition is also consistent with the definition of partially coherent phase used in the literature (Eqn. \ref{eqn:WignerPhase2}), and will therefore be used in what follows for the expected polarization angle gradient of our vector beam. 

\subsection{Conservation of the generalized Stokes parameters \label{sec:conservation}}

To conclude this section, we point out an interesting result concerning the conservation of the generalized  Stokes parameters during propagation through a (possibly) inhomogeneous medium.  Since each of the parameters $s_\nu$  obey the same transport equation (\ref{eqn:vec_cont}), the integrals of these quantities in the transverse plane are constant and are therefore conserved during propagation.  
To see that this is true, expand Eq. (\ref{eqn:vec_cont}), recall the definition of the material derivative $D(\cdot)/Dz\equiv \partial_z(\cdot)+(\vec{v}\cdot\nabla_X)(\cdot)$, and then convert to Lagrangian coordinates whereby the spatial coordinates $\vec{x}$ are written as functions of $z$ and the initial value, that is, $\vec{x}\rightarrow \vec{x}_z(\vec{x}_0)$ (which we will abbreviate as simply $\vec{x}_z$ and the corresponding gradient $\nabla_{X_z}$).  These steps can be written
\begin{align}
    \partial_z\rho_\nu(\vec{x},z,\omega)+\nabla_X \rho_\nu(\vec{x},z,\omega)\cdot\vec{v}_\nu(\vec{x},z,\omega)+\rho_\nu(\vec{x},z,\omega)\left[\nabla_X\cdot\vec{v}_\nu(\vec{x},z,\omega)\right]&=0\nonumber \\
    \frac{D\rho_\nu(\vec{x},z,\omega)}{Dz}+\rho_\nu(\vec{x},z,\omega)\left[\nabla_X\cdot\vec{v}_\nu(\vec{x},z,\omega)\right]&=0\nonumber \\
    \frac{d\rho_\nu(\vec{x}_z,\omega)}{dz}+\rho_\nu(\vec{x}_z,\omega)\left[\nabla_{X_z}\cdot\vec{v}_\nu(\vec{x}_z,\omega)\right]&=0,
    \hspace{.5cm}~\nu=0,1,2,3.
    \label{eqn:vec_cont_exp}
\end{align}
The resulting {\it ordinary
} differential equation possesses the solution
\begin{align}
    \boldsymbol{\rho}(\vec{x}_z,\omega)&=\boldsymbol{\rho}(\vec{x}_0,\omega)\exp{\left(-\int_{s=0}^z\nabla_{X_s}\cdot\vec{\bf v}(\vec{x}_s,\omega)d\vec{x}_s\right)}.
    \label{eqn:divform}
\end{align}
Therefore, the integral of the generalized Stokes parameters at any point along the propagation path is related to their initial values via the divergence of the velocity field. It can also be shown (see Appendix \ref{sec:Jacobian}, ref. \cite{Constantin:08}) that the expression (\ref{eqn:divform}) can be written alternatively  as
\begin{align}
    \boldsymbol{\rho}(\vec{x}_z,\omega)&=\det|J_{\vec{x}_0}(\vec{x}_z)|^{-1}\boldsymbol{\rho}(\vec{x}_0,\omega).
    \label{eqn:Jacobian}
\end{align}
where the Jacobian $J_{\vec{x}_0}(\vec{x}_z)$ is the derivative of the coordinate functions $\vec{x}_z(\vec{x}_0,z)$ with respect to the fixed (initial) coordinates $\vec{x}_0$.  If the Lagrangian coordinate mappings $\vec{x}_z(\vec{x}_0,z)$ are one-to-one, smooth functions of $\vec{x}_0$, this expression can be shown via the change of variables theorem to be equivalent to the integral formulation \cite{Brenier:00}
\begin{align}
    \int_X \boldsymbol{\rho}(\vec{x}_z,\omega)d\vec{x}_z&=\int_{X}\boldsymbol{\rho}(\vec{x}_0,\omega)d\vec{x}_0
    \label{eqn:integralForm}
\end{align}
(see also Villani Chapter 11 \cite{Villani:08}).
Thus, it can be stated that the {\it integral of each of the generalized Stokes parameters over the transverse plane is conserved on propagation}.  Interestingly, this conclusion was reached by an entirely different approach in \cite{Korotkova:08}.  However, in that work the claim was made with respect to propagation in free-space whereas here we see the integrals of the Stokes parameters are conserved even in an inhomogeneous medium. 

\section{Momentum Equation for Partially-Coherent, Linearly-Polarized Light \label{sec:momentum}}

The above discussion defined both intensity and velocity in the general case of a propagating, linearly polarized, partially coherent optical field.  Each of the generalized Stokes parameters were shown to be transported with a single velocity.  In this section, we will derive the equation that governs this velocity.   As we described in section (\ref{sec:transport}), the process for obtaining the continuity equation can be viewed as taking the zeroth ``moment'' of the fundamental transport equation with respect to wavenumber, that is, integrating \ref{eqn:parabolic7}; (see again \cite{Marcuvitz:80}, \cite{Degond:03}, and \cite{Jungel:06}).  Thus, to derive the partially coherent momentum equation we multiply (\ref{eqn:parabolic7}) by $\vec{\xi}$ and integrate over wavenumber. The result is written
\begin{align}
\partial_z \int_{\R^2} \vec{\xi} \widehat{\mathbfcal{S}}(\vec{x},\vec{\xi},z,\omega) 
d{\vec{\xi}}+\nabla_X\cdot \frac{1}{k_0}\int_{\R^2} (\vec{\xi}\otimes \vec{\xi})\widehat{\mathbfcal{S}}(\vec{x},\vec{\xi},z,\omega)d\vec{\xi}=\int_{\R^2}\int_{\R^2} ik_0 \vec{\xi}~\widehat{\mathbfcal{S}}(\vec{x},\vec{\xi}',z,\omega)\Phi(\vec{x},\vec{\xi}-\vec{\xi}',z)d\vec{\xi}'d\vec{\xi}.
\label{eqn:moment2}
\end{align}

The first challenge associated with (\ref{eqn:moment2}) is the term on the right-hand-side, integration of Wigner potential.  This term can be greatly simplified if we are willing to restrict the model to a {\it weakly} inhomogeneous medium where the transverse gradient in dielectric constant is small at a given transverse location $\vec{x}$.  In this case, we explore a series solution for the terms inside the potential function, for example,
\begin{align}
    \epsilon_1(\vec{x}+\frac{\vec{x}'}{2},z)-\epsilon_1(\vec{x}-\frac{\vec{x}'}{2},z)\approx \epsilon_1(\vec{x},z)\pm\nabla_{X}\epsilon_1(x,z)\cdot \vec{x}'+\cdots
\end{align}
so that the potential functions (\ref{eqn:vecPot},\ref{eqn:Markov}) become
\begin{align}
\Phi(\vec{x},\vec{x}',z)\approx \vec{x}'\cdot \nabla_X \left[\frac{1}{2}\epsilon_1(\vec{x},z)\right]+O(\vec{x'}^3)\nonumber \\
\Phi(\vec{x},\vec{x}',z)\approx \vec{x}'\cdot \nabla_X\left[-\frac{ik_0}{4}A(0,z)\right]+O(\vec{x'}^2)
    \label{eqn:expansions}
\end{align}
%or in the case of the stochastic Wigner potential
%\begin{align}
%\Phi(\vec{x},\vec{x}',z)&\approx \frac{1}{2}\vec{x}'\cdot \nabla_X A(\vec{x},z).
%\end{align}
%We will continue the derivation in the deterministic setting for $n^2(\vec{x},z)$ noting simply that this term is the deterministic limit of $A(\vec{x},z)=\langle %n(\vec{x},z)n(\vec{x},z)\rangle$.  
Using these expressions, and representing the terms inside the transverse gradient generically as $g(\vec{x})$ (to accommodate a deterministic or stochastic medium), we we see that the inner double integral on the right hand side of Eq. (\ref{eqn:moment2}) can be expanded using the Fourier representation of the potential as
\begin{align}
&\left(\frac{1}{2\pi}\right)^2\int_{\R^2}\int_{\R^2} ik_0 \widehat{\mathbfcal{S}}(\vec{x},\vec{\xi}',z,\omega) \left[e^{i(\vec{\xi}-\vec{\xi}')\cdot\vec{x}'}\vec{x}'\cdot \nabla_X g(\vec{x},z)d\vec{x'}\right]d\vec{\xi}'\nonumber \\
&=-\left(\frac{1}{2\pi}\right)^2\int_{\R^2}\int_{\R^2} k_0 \widehat{\mathbfcal{S}}(\vec{x},\vec{\xi}',z,\omega) \left[\nabla_{\xi'}e^{i(\vec{\xi}-\vec{\xi}')\cdot\vec{x}'}\cdot \nabla_X g(\vec{x},z)d\vec{x'}\right]d\vec{\xi}'\nonumber \\
&=-\left(\frac{1}{2\pi}\right)^2\int_{\R^2}k_0\widehat{\mathbfcal{S}}(\vec{x},\vec{\xi}',z,\omega) \nabla_{\xi'}\left[\int_{\R^2} e^{i(\vec{\xi}-\vec{\xi}')\cdot\vec{x}'}d\vec{x}'\right]\cdot \nabla_X g(\vec{x},z)d\vec{\xi}'
\nonumber \\
&=-\int_{\R^2}k_0\nabla_{\vec{\xi}'}~\delta(\vec{\xi}-\vec{\xi}')\widehat{\mathbfcal{S}}(\vec{x},\vec{\xi}',z,\omega)\cdot\nabla_Xg(\vec{x},z)d\vec{\xi}'\nonumber \\
&=-\int_{\R^2}k_0\delta(\vec{\xi}-\vec{\xi}')\nabla_{\vec{\xi'}}~\widehat{\mathbfcal{S}}(\vec{x},\vec{\xi}',z,\omega)\cdot\nabla_Xg(\vec{x},z)d\vec{\xi}'\nonumber \\
&=-k_0\nabla_{\vec{\xi}}~\widehat{\mathbfcal{S}}(\vec{x},\vec{\xi},z,\omega)\cdot\nabla_X g(\vec{x},z)
\label{eqn:linearPot}
\end{align}
where we have used the trick found in \cite{Nedjalkov:11}( Eq. 5.29), which is to note that $\nabla_{\xi'}\left(e^{i(\vec{\xi}-\vec{\xi}')\cdot\vec{x}'}\right)=-i\vec{x}'e^{i(\vec{\xi}-\vec{\xi}')\cdot\vec{x}'}$. The expression (\ref{eqn:linearPot}) is the ``linearized'' Wigner potential (see, for example, \cite{Yin:13}) and transforms Eqn. (\ref{eqn:moment2}) into
\begin{align}
\partial_z \int_{\R^2} \vec{\xi} \widehat{\mathbfcal{S}}(\vec{x},\vec{\xi},z,\omega) 
d{\vec{\xi}}+\nabla_X\cdot \frac{1}{k_0}\int_{\R^2} (\vec{\xi}\otimes \vec{\xi})\widehat{\mathbfcal{S}}(\vec{x},\vec{\xi},z,\omega)d\vec{\xi}=-k_0\nabla_Xg(\vec{x},z)\int_{\R^2} \left(\vec{\xi}\cdot\nabla_{\vec{\xi}}~\widehat{\mathbfcal{S}}(\vec{x},\vec{\xi},z,\omega)\right)d\vec{\xi}.
\label{eqn:moment3}
\end{align}

We have already shown via Eq. (\ref{eqn:velnum}) that the first term in (\ref{eqn:moment3}) can be written
\begin{align}
   \partial_z \int_{\R^2} \vec{\xi~} \widehat{s'_\nu}(\vec{x},\vec{\xi},z,\omega) 
d{\vec{\xi}}&=k_0\partial_z \left[{ \rho}_\nu(\vec{x},z,\omega)\vec{v}_0(\vec{x},z,\omega)\right],~\nu=0,\cdots,2 
\end{align}
while for the third Stokes parameter we showed previously in Eqn. (\ref{eqn:s3moment}) that
\begin{align}
\partial_z \int_{\R^2} \vec{\xi}~\widehat{s'_3}(\vec{x},\vec{\xi},z,\omega)
d{\vec{\xi}}&=-k_0\partial_z\left[\rho_0(\vec{x},z,\omega)\vec{\Omega}(\vec{x},z,\omega)\right].
\end{align}
The integral in the last term can also be simplified via integration by parts.  Presuming that the Stokes parameters vanish at the boundaries of integration (as $|\vec{\xi}|\rightarrow \infty$), the result is simply
\begin{align}
    \int_{\R^2} \left(\vec{\xi}\cdot\nabla_{\vec{\xi}}~\widehat{\mathbfcal{S}}(\vec{x},\vec{\xi},z,\omega)\right)d\vec{\xi}=-{\boldsymbol\rho}(\vec{x},z,\omega)
\end{align}
for the first three Stokes parameters ($\nu=0,1,2$), while the integral disappears for the third since, in the absence of ellipticity, $\rho_3(\vec{x},z,\omega)=0$. 
With these simplifications, Eqn. (\ref{eqn:moment3}) becomes
\begin{align}
\partial_z \left[{\rho}_\nu(\vec{x},z,\omega)\vec{v}_0(\vec{x},z,\omega)\right]+\nabla_X\cdot \frac{1}{k_0^2}\int_{\R^2} (\vec{\xi}\otimes \vec{\xi})\widehat{s'_\nu}(\vec{x},\vec{\xi},z,\omega)d\vec{\xi}&= \rho_\nu(\vec{x},z,\omega)\nabla_X g(\vec{x},z),~\nu=0,1,2\nonumber \\
-\partial_z\left[\rho_0(\vec{x},z,\omega)\vec{\Omega}(\vec{x},z,\omega)\right]+\nabla_X\cdot \frac{1}{k_0^2}\int_{\R^2} (\vec{\xi}\otimes \vec{\xi}) \widehat{s'_3}(\vec{x},\vec{\xi},z,\omega)d\vec{\xi}&=0
\label{eqn:moment4}
\end{align}
Equation (\ref{eqn:moment4}) is a vector equation describing the evolution of the ``momentum'' ${\boldsymbol\rho}\circ\vec{\bf v}$ for each of the generalized Stokes parameters and it allows us to develop expressions for the unknowns $\vec{v}_0$ and $\vec{\Omega}$.  As we noted in section (\ref{sec:velocity}), because there is only a single velocity associated with the first 3 equations, we need only focus on the expression (\ref{eqn:moment4}) for $\nu=0$ and the last expression where $\nu=3$.  We also note that the right hand side of the second expression is zero for the linearized potential since $\rho_3(\vec{x},z,\omega)=0$.  As we will show, this leads to the conclusion that the polarization angle gradient will not change on propagation.  Depolarization {\it during} propagation would therefore be a consequence of higher-order terms in the expansion of the Wigner potential or retention of the electric field divergence in the starting wave equation (see e.g., \cite{Charnotskii:16}).  

Recall the continuity equation (\ref{eqn:vec_cont}) was obtained as the zeroth moment of the wave-vector with respect to the Stokes parameters (Eqn. \ref{eqn:parabolic7}) and contained the ``velocity'', which is the first moment of the wave-vector with respect to the Stokes parameters.  In turn, (\ref{eqn:moment4}) contains the {\it second} moment of the wave-vector with respect to the Stokes parameters.  As might be expected, if we were to construct the expression for the second moment it would contain the integral of third order wave-vector terms and so on.  This is the so-called ``closure problem'' (see \cite{Jungel:06} Eqs. 3.3-3.5 and surrounding discussion of  \cite{Manfredi:21}).  

Obtaining closure requires an expression for this second moment (the term involving the outer product) in terms of already defined quantities.  In our context, this means writing the integral involving the outer product in terms of $\rho$ and $\vec{v}$.  This can be accomplished by again leveraging our electric field model (\ref{eqn:modelstruct}).
Note that using this particular field model as an expression to fix the closure problem was employed in both  \cite{Marcuvitz:80,Manfredi:21}.  
To see how this can be done,  expand the second term in (\ref{eqn:moment4}) as was done in \cite{Marcuvitz:80}.  Doing so requires us to expand the Wigner-averaged outer product into its four constituent terms which are of the two basic forms.  

Using the first of equations (\ref{eqn:moment4}) as an example, the outer product is seen to produce the following four integrals,
\begin{align}
    \int_{\R^2}\xi_{ii}^2&\left[\widehat{W}_{XX}(\vec{x},\vec{\xi},z,\omega)+\widehat{W}_{YY}(\vec{x},\vec{\xi},z,\omega)\right]d\vec{\xi},
    \hspace{0.5cm}
    i\in X,Y\nonumber \\
    \int_{\R^2}\xi_i\xi_j&\left[\widehat{W}_{XX}(\vec{x},\vec{\xi},z,\omega)+\widehat{W}_{YY}(\vec{x},\vec{\xi},z,\omega)\right]d\vec{\xi},
    \hspace{0.5cm}i,j\in X,Y.
    \label{eqn:Wigmoments}
\end{align}
Both the unpolarized and polarized portions of the mixture model can be simplified in the same manner as was done for the velocity terms, Eq. (\ref{eqn:veldef}). Substituting in the definition of either the unpolarized/polarized Wigner Transforms $\widehat{W}_{XX}(\vec{x},\vec{\xi},z,\omega)$ and $\widehat{W}_{YY}(\vec{x},\vec{\xi},z,\omega)$, the first of the needed moments in (\ref{eqn:Wigmoments}) can be simplified as
\begin{align}
\int_{\R^2}\xi_{XX}^2&\left[\widehat{W}_{XX}(\vec{x},\vec{\xi},z,\omega)+\widehat{W}_{YY}(\vec{x},\vec{\xi},z,\omega)\right]d\vec{\xi}
\nonumber \\
%    &=\int_{\R^2}\xi_x^2\int_{\R^2} \left\langle E_X\left(\vec{x}-\frac{\vec{x}'}{2},z,\omega\right)E_X^*\left(\vec{x}+\frac{\vec{x}'}{2},z,\omega\right)+E_Y\left(\vec{x}-\frac{\vec{x}'}{2},z,\omega\right)E_Y^*\left(\vec{x}+\frac{\vec{x}'}{2},z,\omega\right)\right\rangle e^{-i\vec{\xi}\cdot\vec{x}'}d\vec{x}'d\vec{\xi}\nonumber \\
%    &=\int_{\R^2}\left\langle E_X\left(\vec{x}-\frac{\vec{x}'}{2},z,\omega\right)E_X^*\left(\vec{x}+\frac{\vec{x}'}{2},z,\omega\right)+E_Y\left(\vec{x}-\frac{\vec{x}'}{2},z,\omega\right)E_Y^*\left(\vec{x}+\frac{\vec{x}'}{2},z,\omega\right)\right\rangle\int \xi_x^2 e^{-i\xi_x x'}d\xi_x\int e^{-i\xi_y y'}d\xi_yd\vec{x}'\nonumber \\
%    &=-\int_{\R^2} \left\langle E_X\left(\vec{x}-\frac{\vec{x}'}{2},z,\omega\right)E_X^*\left(\vec{x}+\frac{\vec{x}'}{2},z,\omega\right)+E_Y\left(\vec{x}-\frac{\vec{x}'}{2},z,\omega\right)E_Y^*\left(\vec{x}+\frac{\vec{x}'}{2},z,\omega\right)\right\rangle\delta''(x')\delta(y')d\vec{x}'\nonumber \\
    &=-\frac{\partial^2}{\partial x'^2}\left\langle E_X\left(\vec{x}-\frac{\vec{x}'}{2},z,\omega\right)E_X^*\left(\vec{x}+\frac{\vec{x}'}{2},z,\omega\right)+E_Y\left(\vec{x}-\frac{\vec{x}'}{2},z,\omega\right)E_Y^*\left(\vec{x}+\frac{\vec{x}'}{2},z,\omega\right)\right\rangle\Bigg|_{\begin{array}{c}x'\rightarrow 0\\ y'\rightarrow 0\end{array}}.
\end{align}
%In the final line we have made use of the identity Eq. (7) of \cite{Boykin:03} concerning integrals over derivatives of the delta function.  
Using our model for the electric field, Eq. (\ref{eqn:modelstruct}), substituting into the above and taking the  derivative and subsequent limits (and noting the procedure in the ``y'' direction is the same) yields
\begin{align}
 \int_{\R^2}\xi_{XX}^2\left[\widehat{W}_{XX}(\vec{x},\vec{\xi},z,\omega)+\widehat{W}_{YY}(\vec{x},\vec{\xi},z,\omega)\right]d\vec{\xi}&=k_0^2\rho_0(v_x^2+\Omega_x^2)+\frac{(\partial_x\rho_0)^2}{4\rho_0}-\frac{\partial_x^2\rho_0}{4}\nonumber \\
\int_{\R^2}\xi_{YY}^2\left[\widehat{W}_{XX}(\vec{x},\vec{\xi},z,\omega)+\widehat{W}_{YY}(\vec{x},\vec{\xi},z,\omega)\right]d\vec{\xi}&=k_0^2\rho_0(v_y^2+\Omega_y^2)+\frac{(\partial_y\rho_0)^2}{4\rho_0}-\frac{\partial_y^2\rho_0}{4}
\end{align}
where $v_{x,y}$ and $\Omega_{x,y}$ are, respectively, the ``$x$'' and ``$y$'' components of those vectors.  In this and several subsequent expressions we omit the arguments $(\vec{x},z,\omega)$ from the model quanitities for brevity.  The ``mixed'' averages follow a similar procedure and are found to be
\begin{align}
    \int_{\R^2}&\xi_X\xi_Y\left[\widehat{W}_{XX}(\vec{x},\vec{\xi},z,\omega)+\widehat{W}_{YY}(\vec{x},\vec{\xi},z,\omega)\right]d\vec{\xi}=
    k_0^2\rho_0(v_xv_y+\Omega_x\Omega_y)+\frac{\partial_x\rho_0\partial_y\rho_0}{4\rho_0}-\frac{\partial_{xy}\rho_0}{4}
\end{align}
so that we may finally write
\begin{align}
    \frac{1}{k_0^2}\int_{\R^2} \left(\vec{\xi}\otimes\vec{\xi}\right)&\left[\widehat{W}_{XX}(\vec{x},\vec{\xi},z,\omega)+\widehat{W}_{YY}(\vec{x},\vec{\xi},z,\omega)\right]d\vec{\xi}=\nonumber \\
    &\rho_0\vec{v}\otimes\vec{v}+\rho_0\vec{\Omega}\otimes\vec{\Omega}+\frac{1}{4k_0^2}\left[\begin{array}{cc} \frac{(\partial_x\rho_0)^2}{\rho_0^2}-\frac{\partial_x^2\rho_0}{\rho_0} & \hspace{0.5cm} \frac{\partial_x\rho_0\partial_y\rho_0}{\rho_0^2}-\frac{\partial_{xy}\rho_0}{\rho_0}\\
    \frac{\partial_x\rho_0\partial_y\rho_0}{\rho_0^2}-\frac{\partial_{xy}\rho_0}{\rho_0} & \hspace{0.5cm}\frac{(\partial_y\rho_0)^2}{\rho_0^2}-\frac{\partial_y^2\rho_0}{\rho_0}
    \end{array}\right]\rho_0
\end{align}
Repeating this same process for the third Stokes parameter (second equation in (\ref{eqn:moment4})) yields the equations
\begin{align}
\partial_z\left(\begin{array}{c} \rho_0\vec{v}\\ -\rho_0\vec{\Omega}\end{array}\right)
&+\left(\begin{array}{c} \nabla_X\cdot\left[\rho_0\vec{v}\otimes \vec{v}+\rho_0R+\rho_0\vec{\Omega}\otimes\vec{\Omega}\right]\\
-\nabla_X\cdot\left[\rho_0\vec{v}\otimes\vec{\Omega}+\rho_0\vec{\Omega}\otimes \vec{v}\right]\end{array}\right)=\left(\begin{array}{c}\rho_0\nabla_Xg(\vec{x},z)\\ 0\end{array}\right)
\label{eqn:veceikonal}
\end{align}
where we have defined the matrix
\begin{align}
R&=\frac{1}{4k_0^2}\left[\begin{array}{cc} \frac{(\partial_x\rho_0)^2}{\rho_0^2}-\frac{\partial_x^2\rho_0}{\rho_0} & \hspace{0.6cm} \frac{\partial_x\rho_0\partial_y\rho_0}{\rho_0^2}-\frac{\partial_{xy}\rho_0}{\rho_0}\\
    \frac{\partial_x\rho_0\partial_y\rho_0}{\rho_0^2}-\frac{\partial_{xy}\rho_0}{\rho_0} & \hspace{0.6cm}
\frac{(\partial_y\rho_0)^2}{\rho_0^2}-\frac{\partial_y^2\rho_0}{\rho_0}
    \end{array}\right]
\end{align}
A convenient simplification occurs upon expansion of the outer product via $\nabla\cdot(\vec{B}\otimes\vec{A})=\vec{A}(\nabla\cdot\vec{B})+(\vec{B}\cdot\nabla)\vec{A}
$.  This identity, along with Eqs. (\ref{eqn:transport2}) and (\ref{eqn:divgrad}), specifically $\partial_z\rho_0+\nabla_X\cdot\left(\rho_0\vec{v}\right)=0$ and $\nabla_X\cdot(\rho_0\mathcal{P}\vec{\Omega})=0$, transforms  (\ref{eqn:veceikonal}) into
\begin{align}
\left(\begin{array}{c}\rho_0D\vec{v}/Dz\\
%\rho_1 D\vec{v}/Dz\\
%\rho_2 D\vec{v}/Dz\\
\rho_0D\vec{\Omega}/Dz
\end{array}
\right)&+
\left(\begin{array}{c} \nabla_X\cdot\left[\rho_0R+\rho_0\vec{\Omega}\otimes\vec{\Omega}\right]\\
%\nabla_X\cdot\left[\rho_1 R+\Gamma\rho_2\right]\\
%\nabla_X\cdot\left[\rho_2 R-\Gamma\rho_1\right]\\
\left(\rho_0\vec{\Omega}\cdot\nabla_X\right)\vec{v}\end{array}\right)=\left(\begin{array}{c}\rho_0\nabla_Xg(\vec{x},z)\\
%\rho_1\nabla_Xn^2\\
%\rho_2\nabla_Xn^2\\
0
\end{array}
\right).
\label{eqn:veceikonal2}
\end{align}
where we have again leveraged the material derivative defined earlier in section (\ref{sec:conservation}).  The only components of (\ref{eqn:veceikonal2}) that are influenced by the DOP are those involving the polarization angle gradient. To simplify further, note that the divergence of $\rho_0(\vec{x},z,\omega) R(\vec{x},z,\omega)$ is
\begin{align}
\nabla_X\cdot \left[\rho_0(\vec{x},z,\omega) R(\vec{x},z,\omega)\right]&=-\frac{1}{2k_0^2}\rho_0(\vec{x},z,\omega)\nabla_X\left(\frac{\nabla_X^2\rho_0^{1/2}(\vec{x},z,\omega)}{\rho_0^{1/2}(\vec{x},z,\omega)}\right).
\end{align}
Expanding the outer product and invoking  (\ref{eqn:divgrad}), the first of Eqns (\ref{eqn:veceikonal2}) can be written
\begin{align}
    \frac{D\vec{v}_0(\vec{x},z,\omega)}{Dz}&=-\left(\vec{\Omega}(\vec{x},z,\omega)\cdot\nabla_X\right)\vec{\Omega}(\vec{x},z,\omega)+\nabla_X g(\vec{x},z)+\frac{1}{2k_0^2}\nabla_X\left(\frac{\nabla_X^2\rho_0^{1/2}(\vec{x},z,\omega)}{\rho_0^{1/2}(\vec{x},z,\omega)}\right).
    \label{eqn:momentum}
\end{align}
which is of precisely the same form as the coherent momentum equation \cite{Nichols:22}.  In the case of a deterministic medium, the gradient potential forcing term is $\nabla_X \epsilon_1/2$.  In \cite{Nichols:22} the commonly used medium assumptions were used whereby, $\langle\epsilon\rangle=1$ and $\epsilon_1=n^2(\vec{x},z)-1$.  In this case Eq. (\ref{eqn:momentum}) matches exactly that found in \cite{Nichols:22} for $\mathcal{P}=1$ (fully polarized light).  For a stochastic, homogeneous medium where $g(\vec{x},z)=-ik_0A(\vec{x},z)/4$ we have
\begin{align}
    \nabla_X g(\vec{x},z)&=-\frac{ik_0}{4}\nabla_X A(0,z)\nonumber \\
    &=-\frac{ik_0}{4}\int_{\R^2} i\vec{\xi} S_{NN}(\vec{\xi},0)e^{i\vec{\xi}\cdot \vec{0}}d\vec{\xi}    \nonumber \\
    &=\frac{k_0}{4}\int_{\R^2} \vec{\xi} S_{NN}(\vec{\xi},0)d\vec{\xi}
\end{align}

Taking as a simple model for the refractive index fluctuations as Eqn. (38) of \cite{Charnotskii:16}, we have that $\nabla_X g(\vec{x},z)=0$, i.e., the {\it gradient} of the refractive index covariance is zero.  Thus, the time-averaged transverse location is unchanged due to the turbulence.  Note, the model (\ref{eqn:momentum}) can also be used to derive known results for ``beam wander'' in the coherent case following the approach in \cite{Nichols:19}.  In that case, an expression for the transverse coordinate variance is developed and is seen to be proportional to the variance of the refractive index gradient (as opposed to the gradient of the variance) and one recovers the established result (see \cite{Nichols:19} for details).

As in our prior work, it is advantageous to switch to Lagrangian coordinates $\vec{x}\rightarrow \vec{x}_z(\vec{x}_0)$ which are functions of propagation distance $z$ and initial value $\vec{x}_0$.  With this transformation, and noting that the ``velocities'' are the coordinate derivatives with respect to $z$, (\ref{eqn:momentum}) becomes
\begin{align}
    \frac{d\vec{v}_0(\vec{x}_z,\omega)}{dz}=\frac{d^2\vec{x}_z(\omega)}{dz^2}&=-\left(\vec{\Omega}(\vec{x}_z,\omega)\cdot\nabla_{X_z}\right)\vec{\Omega}(\vec{x}_z,\omega)+\nabla_{X_z} g(\vec{x}_z)+\frac{1}{2k_0^2}\nabla_{X_z}\left(\frac{\nabla_{X_z}^2\rho_0^{1/2}(\vec{x}_z,\omega)}{\rho_0^{1/2}(\vec{x}_z,\omega)}\right).
    \label{eqn:lagrangemomentum}
\end{align}
This is a sensible model as it suggests that a fully depolarized beam will follow a path that is influenced only by diffraction (last term in \ref{eqn:lagrangemomentum}) and refractive index fluctuations.  However, {\it if} the beam is polarized and {\it if} the polarization angle distribution possesses non-zero first and second spatial derivatives, there is an additional effect that must be considered, given by the first term on the right hand side of (\ref{eqn:momentum}). In the case of diffraction only, the model (\ref{eqn:lagrangemomentum}) recovers the known result for the beam path in both the Gaussian beam \cite{Nichols:19}) and Airy beam \cite{Nichols:22}) situations. 

The second equation in (\ref{eqn:veceikonal2}) governs the evolution of the polarization angle.  Following the derivation in Appendix (\ref{sec:polar}), this expression reduces to
\begin{align}
\frac{D\vec{\Omega}}{Dz}+\left(\vec{\Omega}\cdot\nabla_X\right)\vec{v}=0
\label{eqn:polarang}
\end{align}
or in Lagrangian coordinates
\begin{align}
\frac{d\vec{\Omega}(\vec{x}_z,\omega)}{dz}&=0.
\label{eqn:polarcons}
\end{align}
{\it Thus, for partially coherent light, the time-average polarization gradient as defined in (\ref{eqn:gradangledef2}) will remain unchanged during propagation when observed in Lagrangian coordinates.}  This result is consistent with our prior work which held that in the fully coherent case, the polarization angle remained unchanged on propagation \cite{Nichols:22}.  By definition, depolarization will decrease $\vec{\Omega}$, ultimately resulting in $\vec{\Omega}=0$ for $\mathcal{P}=0$.  However, no mechanism exists in (\ref{eqn:polarcons}) to cause this to occur.  While it is known that the atmosphere can serve as a depolarizing element, under the MA model such an effect can only be observed by retaining higher-order spatial variations in the refractive index (stemming ultimately from retaining the electric field divergence term in the wave equation) \cite{Charnotskii:16,Nichols:19}.  In both of the cited works retention of such terms has almost no influence on beam propagation but for very long distances.  This distance scales as $O(k_0^2\ell_0^3/\sigma_{\epsilon}^2)\sim 10^{10}m$, even when considering strong turbulence (refractive index variance $\sigma_{\eta}^2\sim 10^{-10}$), small inhomogeneities (length scale $\ell_0\sim 10^{-3}m$), and $1.55\mu m$ wavelength light \cite{Charnotskii:16}.  Other works have predicted depolarization will arise due to asymmetries in the initial spatial correlations across the beam face \cite{Wolf:05}.  We have not included this effect in our model given that lasers do not naturally possess such asymmetry and creating such a beam would be challenging. 

In summary, we conclude that whether we are considering coherent, monochromatic, or partially coherent, polychromatic wave propagation, the basic governing equations are the same.  
\begin{align}
    \rho_0(\vec{x}_z,\omega)&=|\det J_{\vec{x}_0}(\vec{x}_z)|^{-1}\rho_0(\vec{x}_0,\omega).\nonumber \\
   \frac{d^2\vec{x}_z}{dz^2}&=-\left(\vec{\Omega}(\vec{x}_z,\omega)\cdot\nabla_{X_z}\right)\vec{\Omega}(\vec{x}_z,\omega)+\nabla_{X_z} g(\vec{x}_z)+\frac{1}{2k_0^2}\nabla_{X_z}\left(\frac{\nabla_{X_z}^2\rho_0^{1/2}(\vec{x}_z,\omega)}{\rho_0^{1/2}(\vec{x}_z,\omega)}\right)
    \label{eqn:momentumFinal}\nonumber \\
    \frac{d\vec{\Omega}(\vec{x}_z,\omega)}{dz}&=0
\end{align}

It is the definition of quantities in the expression that is different in the two cases.  These can be summarized as follows:\\*[0.2in]
{\bf Coherent, monochromatic}
\begin{align}
\rho(\vec{x},z)&=E_X(\vec{x},z)E_X^*(\vec{x},z)+E_Y(\vec{x},z)E_Y^*(\vec{x},z)\nonumber \\
\vec{v}(\vec{x},z)&=\frac{1}{k_0}\nabla_X\phi(\vec{x},z)\nonumber \\
\vec{\Omega}(\vec{x},z)&=\frac{1}{k_0}\nabla_X\gamma(\vec{x},z)
\end{align}
{\bf Partially coherent, polychromatic}
\begin{align}
\rho_0(\vec{x},z,\omega)&=\int_{\R^2} \left[\widehat{W}_{XX}(\vec{x},\xi,z,\omega)+\widehat{W}_{YY}(\vec{x},\xi,z,\omega)\right] d\vec{\xi}=S_{XX}(\vec{x},z,\omega)+S_{YY}(\vec{x},z,\omega)\nonumber \\
    \vec{v}_0(\vec{x},z,\omega)&=
    \frac{1}{k_0}\langle\nabla_X\phi(\vec{x},z,\omega)\rangle=\frac{\int_{\R^2}\vec{\xi}\left[\widehat{W}_{XX}(\vec{x},\vec{\xi},z,\omega)+\widehat{W}_{YY}(\vec{x},\vec{\xi},z,\omega)\right]d\vec{\xi}}{k_0\rho_0(\vec{x},z)}\nonumber \\
    \vec{\Omega}(\vec{x},z,\omega)&=\frac{\mathcal{P}(\vec{x},z,\omega)}{k_0}\langle\nabla_X\gamma(\vec{x},z,\omega)\rangle=\frac{-i\mathcal{P}\int_{\R^2}\vec{\xi}\left[\widehat{W}_{XY}(\vec{x},\vec{\xi},z,\omega)-\widehat{W}_{YX}(\vec{x},\vec{\xi},z,\omega)\right]d\vec{\xi}}{k_0\rho_0(\vec{x},z)}
    \label{eqn:velocitydefs}
\end{align}
where we can see that the latter reduce to the former in the coherent limit.

\section{Examples \label{sec:examples}}
In our previous work we showed that for a particular functional form of polarization angle gradient the beam would ``accelerate'' in the transverse direction following a curved path and we have shown in (\ref{eqn:momentumFinal}) that this is also true in the partially coherent case.  Using a new definition of polarization angle gradient (\ref{eqn:gradangledef2}), we see the the beam will follow the same coherent path as in \cite{Nichols:22} for $\mathcal{P}=1$ but will follow a different path as the DOP decreases $\mathcal{P}<1$.  

Consider first the simple case of a beam with Gaussian intensity distribution of equal amplitude in the $x$ and $y$ components.  As in \cite{Korotkova:04} (see Eqn. 3.6 in the cited work) we take $\mathcal{P}(\vec{x},z)=\mathcal{P}$ as a constant representing the correlation coefficient between the $x$ and $y$ components of the electric field and obeying $0\le \mathcal{P} \le 1$.  In our prior work we set $\mathcal{P}=1$ and used the polarization profile
\begin{align}
\gamma(y_0)&=\frac{\pi}{2}\frac{(y_0-a)^2}{a^2}+\frac{\pi}{8}
\label{eqn:profile}
\end{align}
which varies in $y_0$ only and not in $x_0$.  Thus, the polarization gradient terms becomes simply
\begin{align}
\vec{\Omega}(y_0)&=\left\{0,\mathcal{P}\frac{\pi (y_0-a)}{a^2}\right\}
\end{align}
and the second of Eqs. (\ref{eqn:momentumFinal}) becomes the ordinary differential equations
\begin{align}
\frac{d^2x_z}{dz^2}&=x_0\nonumber \\
\frac{d^2y_z}{dz^2}&=-\frac{\mathcal{P}^2}{k_0^2}\frac{d\gamma(y_0)}{dy_0}\frac{d^2\gamma(y_0)}{dy_0^2}
\label{eqn:diffEq}
\end{align}
subject to the initial velocity $dx_0/dz=dy_0/dz=0$ and initial displacements $x_0$ and $y_0$.  This expression can be solved to give the the partially coherent beam path
\begin{align}
x_z&=x_0\nonumber \\
y_z&=\frac{\mathcal{P}^2\pi^2z^2}{2k_0^2a^4}\left(a-y_0\right)+y_0.
\label{eqn:angledispOld}
\end{align}

\begin{figure}[th]
  \centerline{
   \begin{tabular}{ccc}
    \includegraphics[scale=0.5]{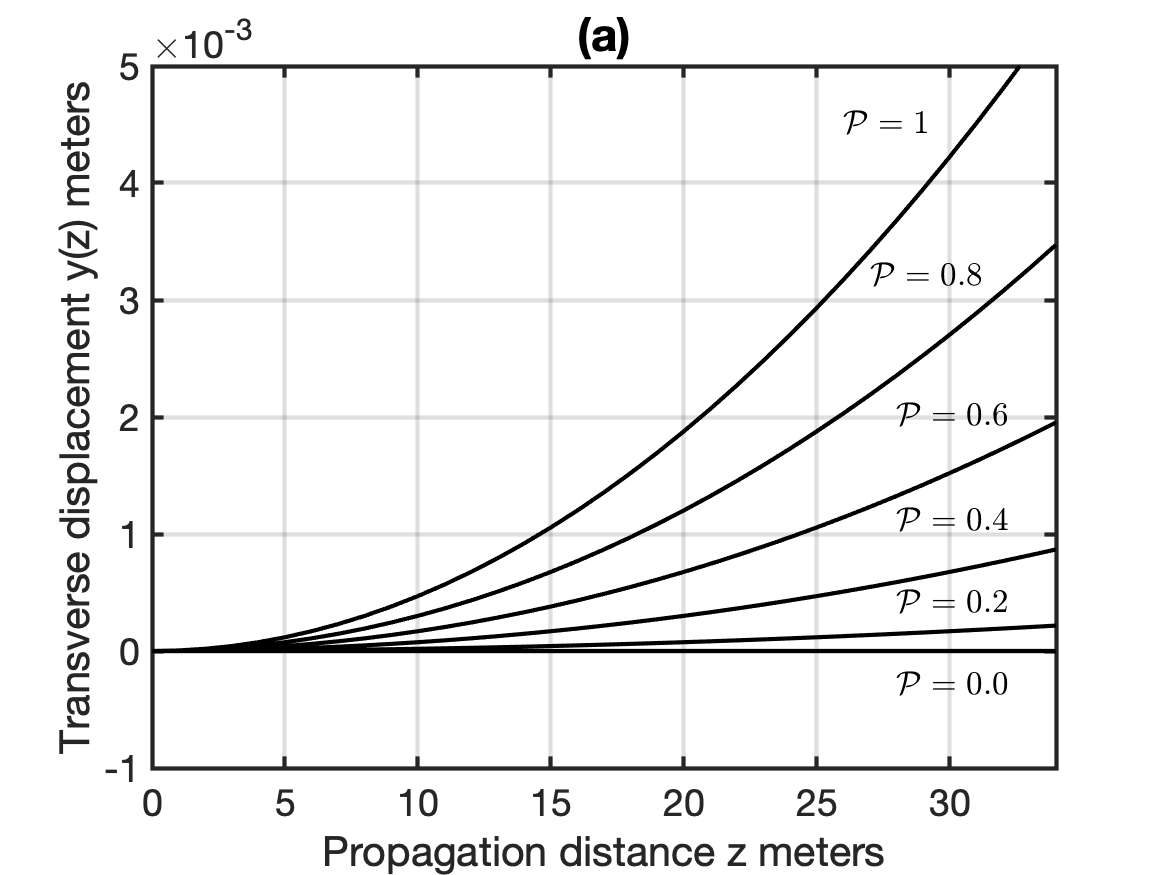} & \hspace*{-0.4in} & 
    \includegraphics[scale=0.5]{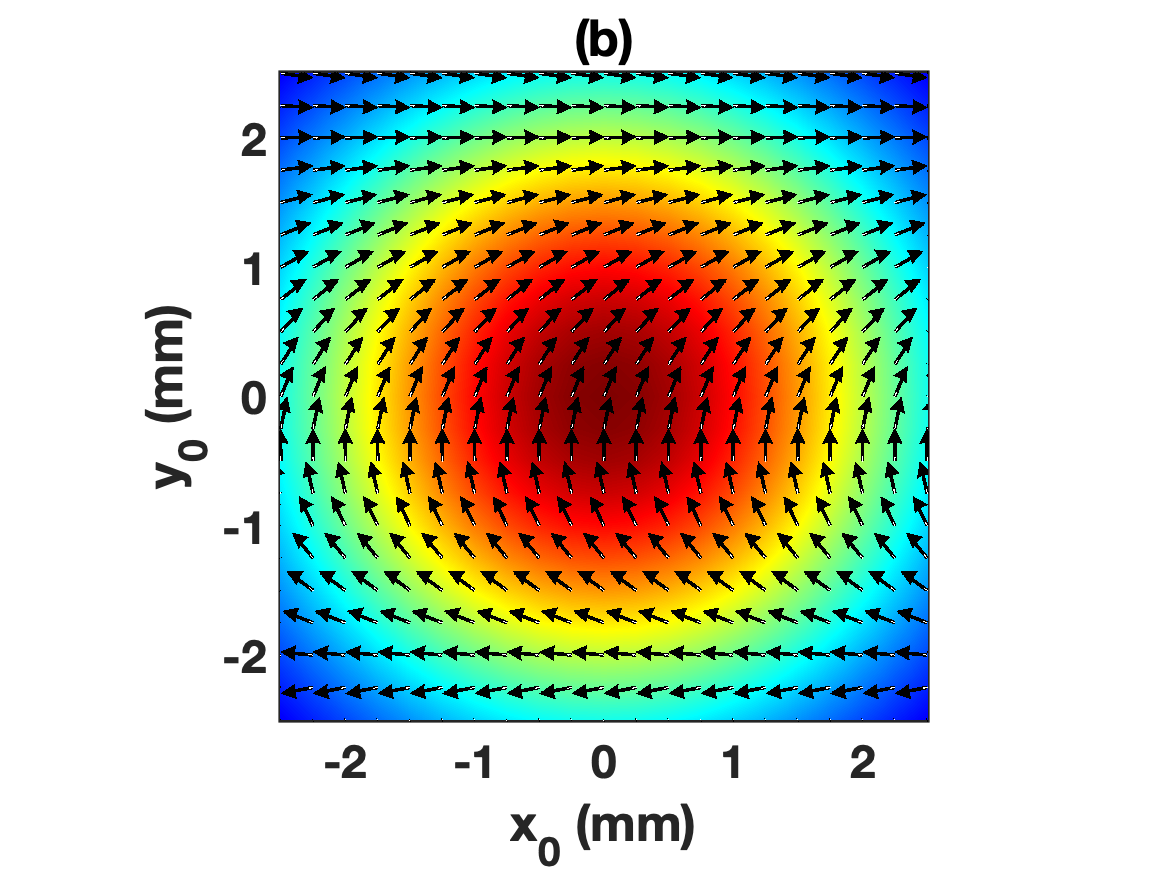}
  \end{tabular}
  }
  \caption{(a) Transverse displacement of the beam center as the DOP is reduced from unity to 0 in increments of 0.2.  The model parameters were the same as those used in our prior work \cite{Nichols:22} and (b) The associated polarization angle profile, Eq. (\ref{eqn:profile})}
  \label{fig:cohereplot}
\end{figure}

The prediction is that for fully polarized light, $\mathcal{P}=1$, the beam will follow the same path obtained in \cite{Nichols:22} while, for unpolarized light, $\mathcal{P}=0$, and the beam will simply follow a straight path.  Figure (\ref{fig:cohereplot}) shows a family of such curves associated with the beam center, $y_0=0$, for different degrees of polarization.  Also shown is the polarization angle profile across the beam face, plotted as a function of the transverse coordinates $\vec{x}_0$.
The result is the expected behavior. As the DOP is reduced, so too is the degree to which the path of the beam center is altered.  We also note that in this example the effects of diffraction could have been included just as they were in our coherent model \cite{Nichols:22}.  The result is the standard Gaussian beam spreading about the path predicted in Eqn. (\ref{eqn:angledispOld}).  

\begin{figure}[th]
  \centerline{
   \begin{tabular}{ccc}
    \includegraphics[scale=0.275]{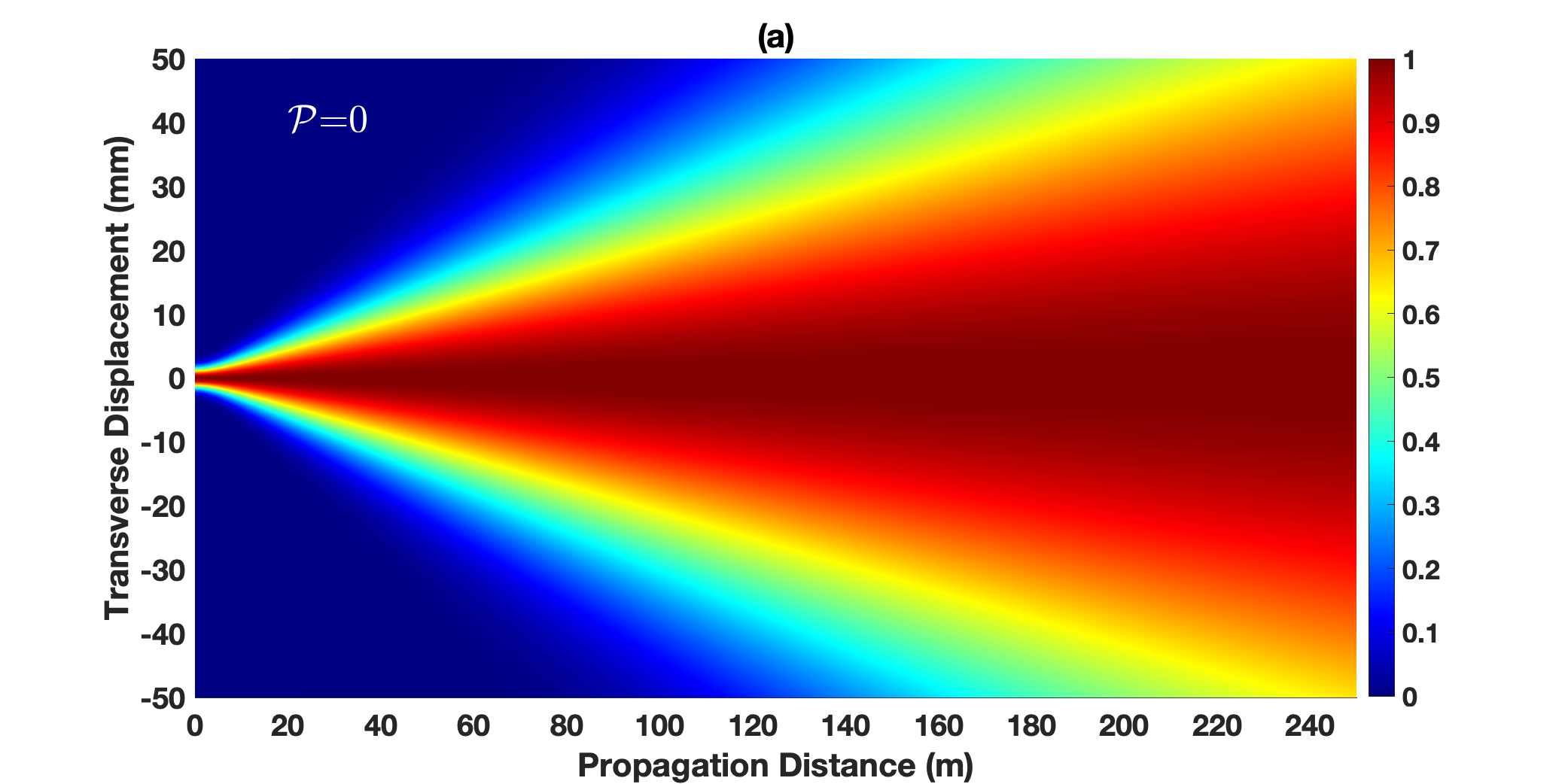} & \hspace*{-0.4in} & 
    \includegraphics[scale=0.275]{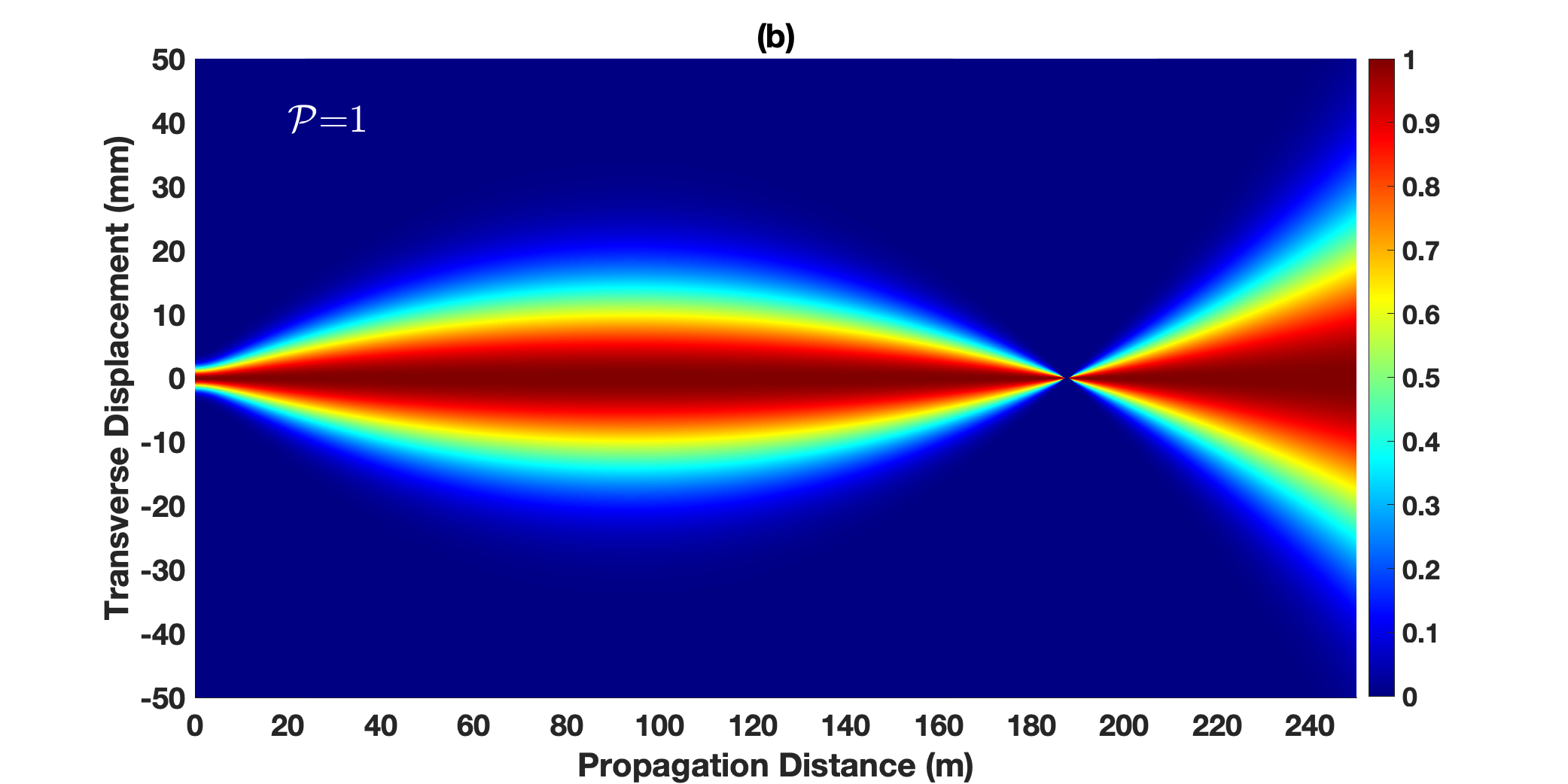}\\
       \includegraphics[scale=0.275]{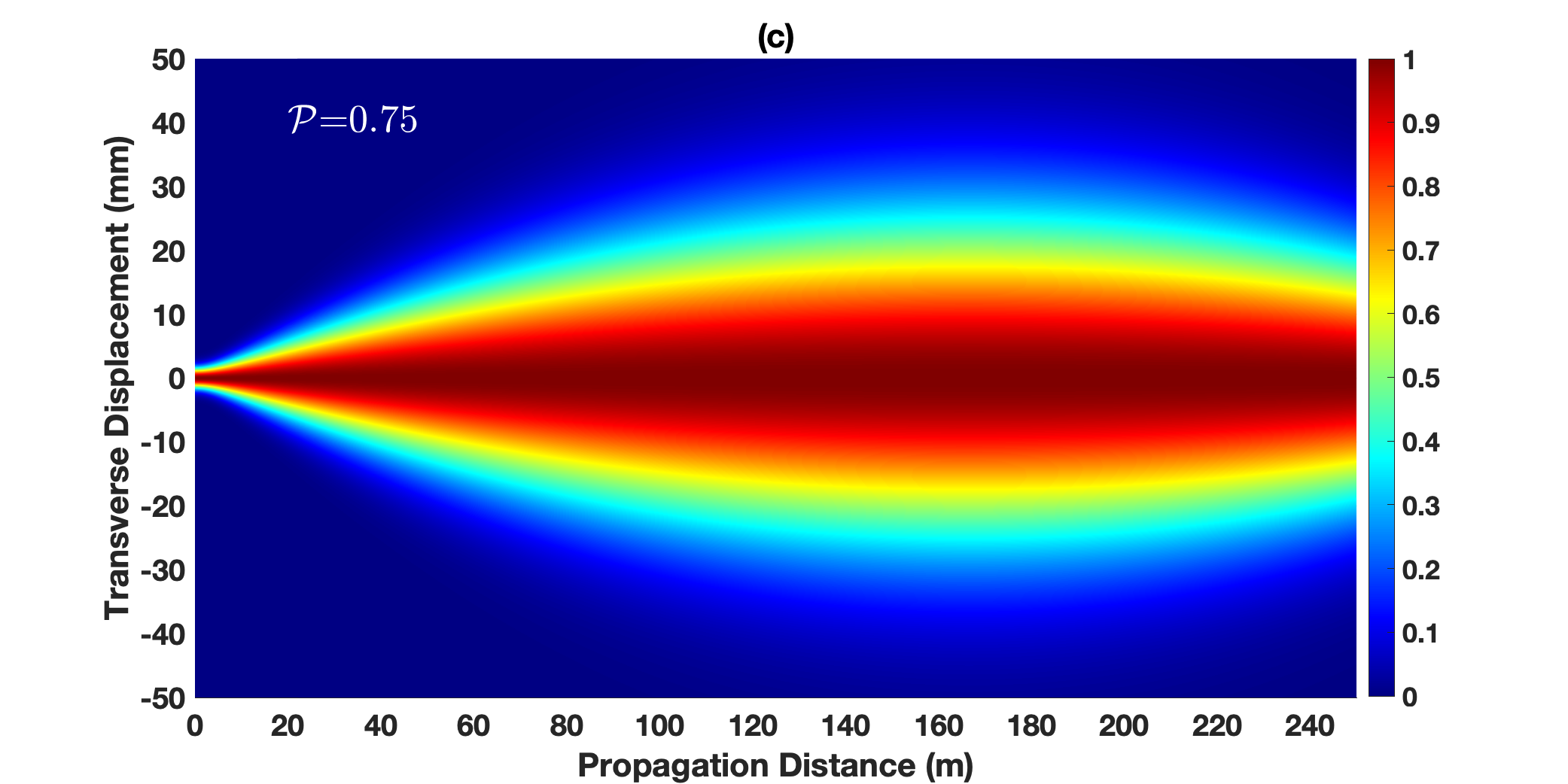} & \hspace*{-0.4in} & 
    \includegraphics[scale=0.275]{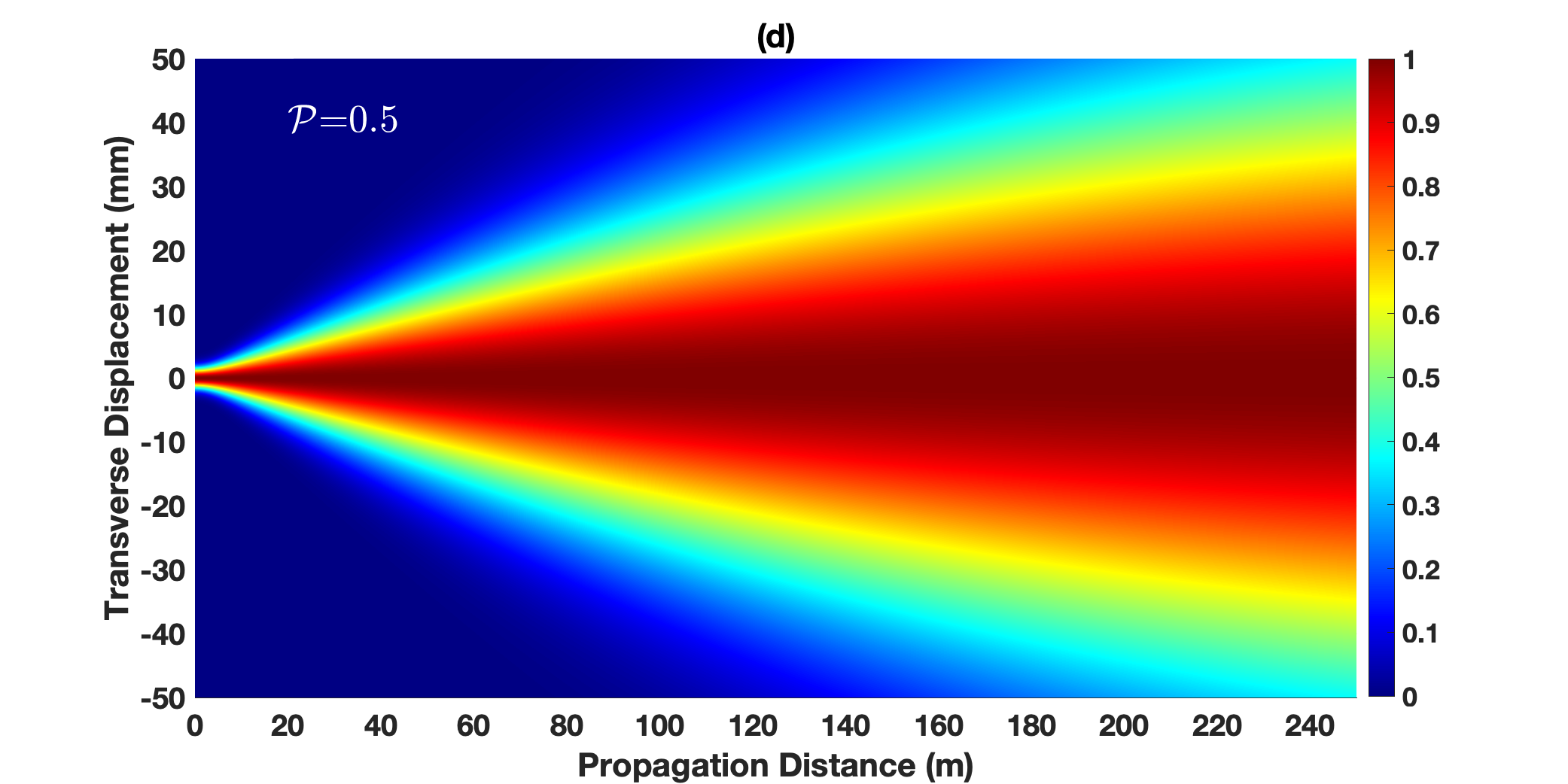}
  \end{tabular}
  }
  \caption{(a) Normal beam diffraction profile over the range $z=[0,250]~m$, (b) diffraction profile associated with the polarization angle distribution given by Eq. (\ref{eqn:symmprofile}) and for $\mathcal{P}=1$.  Here the spreading due to diffraction is compensated by the bending due to the polarization angle gradient. 
 These effects cancel each other out at range $z_c\approx (2a^4k_0)/\left(\mathcal{P}\pi\sigma\right)^2$ (obtained by setting $y_z=0$).  Beyond this range the beam super diffracts and the spreading is proportional to $(z-z_c)^2$.  Plots (c) and (d) use the same polarization profile but with $\mathcal{P}=0.75$ and $\mathcal{P}=0.5$ respectively.}
  \label{fig:DiffractPlot}
\end{figure}

As a second example, we consider the case where both bending and diffraction are retained in the model.  Specifically, we tailor the polarization profile so that the polarization-induced bending acts in opposition to the diffractive effect, thereby mitigating the degree to which the beam spreads on propagation.  Using the polarization profile
\begin{align}
\gamma(y_0)&=\kappa\frac{\pi}{2}\frac{y_0^2}{a^2}-\frac{\pi}{8}
\label{eqn:symmprofile}
\end{align}
we solve the coupled system of Equations (\ref{eqn:momentumFinal}) using an initial Gaussian intensity profile
\begin{align}
\rho_0(\vec{x}_0,\omega)&=I(\omega)e^{-(x_0^2+y_0^2)/(2\sigma^2)}
\end{align}
with width $\sigma$.  The solution is obtained analytically in this case using the approach outlined in 
\cite{Nichols:22} which leads to the transverse path
\begin{align}
x_z&=\frac{x_0\sqrt{z^2+\sigma^4k_0^2}}{\sigma^2k_0}\nonumber \\
y_z&=\frac{-(\mathcal{P}\kappa)^2\pi^2z^2}{2k_0^2a^4}y_0+\frac{y_0\sqrt{z^2+\sigma^4k_0^2}}{\sigma^2k_0} \nonumber \\
&=\left[\left(1+\dfrac{z^2}{k_0^2\sigma^4}\right)^{1/2}-\dfrac{\pi^2}{2}\dfrac{(\mathcal{P}\kappa)^2}{k_0^2a^4}z^2
\right]y_0.
\label{eqn:diffractpath}
\end{align}
The corresponding determinant of the Jacobian is therefore
\begin{align}
\det J_{\vec{x}_0}(\vec{x}_z)&=1+\frac{z^2}{\sigma^4k_0^2}-\frac{(\mathcal{P}\kappa)^2\pi^2z^2\sqrt{z^2+\sigma^4k_0^2}}{2a^4\sigma^2k_0^3}.
\end{align}
and the desired intensity profile on propagation is given by the first of Eqs. (\ref{eqn:momentumFinal}).  The polarization profile (\ref{eqn:symmprofile}) has the effect of bending the outer edges of the beam, $y_0=\pm a/2$, toward the center thus mitigating the outward spreading caused by the diffractive term.  These results can be seen visually in Fig. (\ref{fig:DiffractPlot}) for a fully polarized beam of wavelength $1.55~\mu m$, width $\sigma=1.2~mm$ and extending over transverse spatial dimension $a=4.17~mm$.  In this fully polarized case we set $\kappa=1$.  

The normalized beam intensity is plotted for $z=[0, 250]~m$ in order to clearly show the effects of the polarization gradient.  By altering the polarization of the light across the beam face we can control, to some extent, the degree to which the beam diffracts. Of course, in this example the diffractive effect causes transverse spreading that is linear in $z$ while the bending we show here is quadratic.  Thus, at a distance $z_c$, obtained by setting $y_z=0$ in Eqn. (\ref{eqn:diffractpath}), the rays are predicted to come to a focus. 
 Then
 \begin{align} \label{eq:zc}
 z_c=\dfrac{\sqrt{2}}{\pi^2}
 \dfrac{k_0}{(\mathcal{P}\kappa)^2}
 \left(\dfrac{a^4}{\sigma^2}\right)
 \left[
1+\sqrt{1+\pi^4(\mathcal{P}\kappa)^4
 \left(\dfrac{\sigma}{a}\right)^8}\,\,
 \right]^{1/2}
 \end{align}
 and in the limit $z_c\gg k_0\sigma^2$ we find the particularly simple form  
\begin{align}
%z_c^2&=2\alpha a^4 k_0^2\left(1+\sqrt{1+\frac{\sigma^4}{\alpha a^4}}\right)\nonumber \\
z_c=\frac{2a^4k_0}{\left(\mathcal{P}\pi\sigma\right)^2}.
%\alpha&=\frac{a^4}{\left(P\pi\sigma\right)^4}
\end{align}
(Note also this limit is also obtained when $\pi^4(\mathcal{P}\kappa)^4(\sigma/a)^8<<1$, a condition that will always be met when the beam width $2\sigma$ is less than the output aperture dimension $2a$).
 Beyond $z>z_c$, however, the beam diverges strongly and the intensity rapidly diminishes.  Thus, using this polarization profile diffraction can be mitigated, but only over a prescribed range.  As the beam depolarizes the degree to which diffraction is mitigated will change, however, by increasing the constant $\kappa$ in the polarization gradient such that $\mathcal{P}\kappa=1$, we expect to see the same results as in Fig. (\ref{fig:DiffractPlot}b).  In practice, of course, there will be a limit to how large a polarization gradient can be produced.  As the DOP decreases (with $\kappa=1$) we see the expected results in Figs. (2c-d), namely that the beam gradually moves toward the standard diffraction profile associated with an incoherent beam, Fig. (2a).
 
\begin{figure}%[th]
  \centerline{
\includegraphics[scale=0.48]{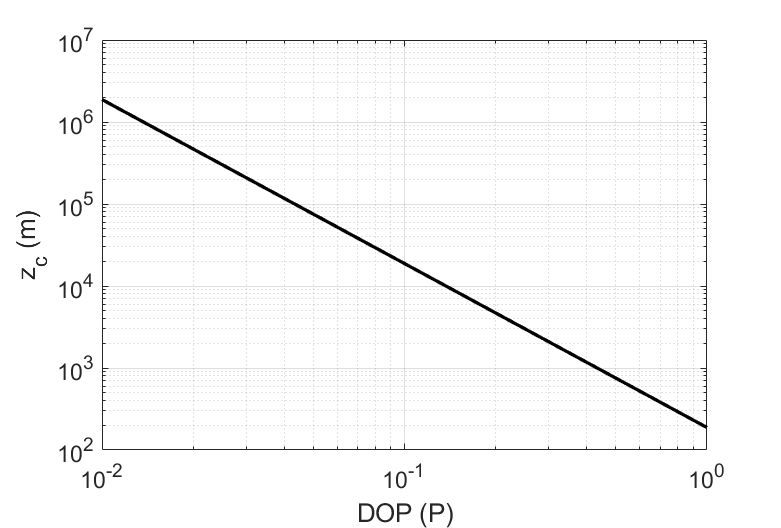}  
  }
  \caption{
  $z_c$ as a function of degree of polarization $\mathcal{P}$ for $\lambda = 1.55~\mu \!m$, $a=4.17$~mm, and $\sigma=1.15$~mm. For these parameters, $z_c\propto\mathcal{P}^{-1}$ and the trend continues indefinitely as $\mathcal{P}\rightarrow 0$.
  }
  \label{fig:zcVsP}
\end{figure}

 Fig. (\ref{fig:zcVsP}) shows a plot of $z_c$ from (\ref{eq:zc}) for fixed polarization gradient $(\kappa=1)$ as a function of degree of polarization $\mathcal{P}$ in the range $0.01\leq\mathcal{P}\leq 1$. The $z_c\propto \mathcal{P}^{-1}$ behavior continues indefinitely as $\mathcal{P}\rightarrow 0$, that is, regardless of the degree of polarization, there exists a $z-$value for which the effects of diffraction could be completely compensated.  Of course in practice, there is a limit to $\kappa$ as one can not generate a beam with an infinitely sharp polarization gradient.  The practical limit will for such a beam will be determined by the resolution of the device used in its creation, e.g., the spatial light modulators used in reference \cite{Nichols:22}.

 The results in Fig.~\ref{fig:zcVsP} are consistent with the following interpretation of the mixture model in (\ref{eqn:breakdown}). The term "mixture" cannot be taken literally to mean that the beam comprises a collection of photons in which a fraction $\mathcal{P}$ are fully polarized and fraction $(1-\mathcal{P})$ are completely unpolarized. If that were true then the polarization gradient could only compensate diffraction for the polarized photons and the beam spread at $z=z_c$ would necessarily increase as $\mathcal{P}$ decreased. Instead, we found that the beam width at the focus, in the $y-$~direction for this example, is independent of $\mathcal{P}$. Hence, we interpret (\ref{eqn:breakdown}) to mean that \emph{every} photon in the beam is described by Stokes vector
\begin{align}
{\bf \widehat{\mathbfcal{S}}}= 
\left(\begin{array}{c} s_0'(\vec{x},\vec{x}',z,\omega)\\
\mathcal{P}s_1'(\vec{x},\vec{x}',z,\omega|\mathcal{P})\\
\mathcal{P}s_2'(\vec{x},\vec{x}',z,\omega|\mathcal{P})\\
\mathcal{P}s_3'(\vec{x},\vec{x}',z,\omega|\mathcal{P})\end{array}\right).
\end{align} 
Every photon thus exhibits properties of partial polarization with degree of polarization $\mathcal{P}$.

\section{Summary}
We have generalized our prior, coherent vector beam model to the partially coherent case by introducing a mixture model for the generalized Stokes vector and by subsequently choosing appropriate definitions of amplitude, phase, and polarization angle gradient in describing propagation of the generalized Stokes parameters.  The resulting ``transport'' model predicts the time-averaged optical path of a vector beam propagating through an inhomogeneous medium (e.g., an atmosphere).  Importantly, the model reduces to that of our previous work on vector beams in the coherent limit \cite{Nichols:22} and matches the predictions found in \cite{Tatarskii:69} and \cite{Charnotskii:16} in the absence of a polarization gradient. 
 Consequently, it predicts the same optical path as in our coherent model in the fully polarized case, while reducing to the standard ``unaltered'' path in the unpolarized situation.
 
 Notably, the model requires a new definition of polarization angle gradient that is consistent with more recent definitions of phase for partially coherent beams.  These definitions become important in the governing equations as the movement of intensity during propagation is governed by phase and angle gradients as opposed to the phase and angle variables.  As a by-product of the model development we were also able to demonstrate conservation of the generalized Stokes parameters on propagation, even in the case where the intervening medium possesses refractive index fluctuations.  This is a generalization of the result established in \cite{Korotkova:08}. 

An obvious prediction of the model Eqn. (\ref{eqn:polarcons}) is that the state of linear polarization (as captured by $\vec{\Omega}$) remains unchanged on propagation from a Lagrangian viewpoint.  Because the definition of $\vec{\Omega}$ includes the DOP, this is also tantamount to the statement that the DOP is unchanged on propagation.  The result is consistent with the result of Charnotskii \cite{Charnotskii:16} which found that no significant depolarization occurs due to the presence of a turbulent atmosphere.  Had we included the higher order refractive index variations (by retaining the electric field divergence in the wave equation) and/or retained the full Wigner potential in forming Eqn. (\ref{eqn:moment3}) (as opposed to using the linearized potential) this effect may have been observable in our model as well.

We considered two such examples, one in which the curved path taken by the beam studied in \cite{Nichols:22} was slowly altered by changing the degree of polarization.  In the coherent case, the path taken is the same as in the cited reference.  As the degree of polarization is reduced so too is degree of beam bending, reducing to propagation in a straight line in the fully incoherent case.  In the second example, we consider both the bending and diffracting beam.  For that example we chose an initial polarization distribution that would mitigate the effects of diffraction over a predictable distance.  The result suggests new class of ``non-diffracting'' beam. 
 The effects of depolarization can be countered in this case by simply increasing the strength of the polarization gradient applied across the transverse dimension of the beam.

In short, the model we have developed can be used to predict vector beam trajectories in a variety of propagation scenarios of interest, including through free-space and a potentially turbulent atmosphere.  As such it provides a powerful tool in forecasting the utility of the newly discovered vector beam bending effect in application.  We note that the model rests on the same two assumptions as were used in the coherent model development \cite{Nichols:22}, namely, the slowly varying envelope approximation and a weakly inhomogeneous medium.

\section{Acknowledgments}

The authors would like to thank the Office of Naval Research Code 33, for support under PE\#0601153N, award \#N0001422WX01660

%As an aside, returning to Eqn. (\ref{eqn:veceikonal2}) we can subtract the second and third equations to yield
%\begin{align}
%\nabla_X\cdot\left[S\rho^{(2)}+\Gamma\rho^{(3)}\right]&=\nabla_X\cdot\left[S\rho^{(3)}-\Gamma\rho^{(2)}\right]
%\end{align}
%which after simplification yields
%\begin{align}
%2k_0\vec{\Omega}\cdot R&=\nabla\cdot(\rho\Gamma)
%\end{align}
%It's unclear to me if this expression provides useful information concerning the propagation problem.  We could, for example, re-write the expression as
%\begin{align}
%\rho\vec{\Omega}&=\frac{1}{2k_0}\left[\nabla_X\cdot(\rho\Gamma)\right]\cdot S^{-1}
%\end{align}

\appendix

\section{Interpretation of Transport Velocity \label{app:velocity}}

In this section we present  a more traditional optics interpretation of the transverse velocity $\vec{v}=k_0^{-1}\nabla_X\phi$ for coherent light. 
In particular, we show how this quantity can be related to the familiar concept of ``optical path length'' (OPL). This allows us to properly view the transverse phase gradient of our electric field as the rate at which intensity is moved in the transverse plane.  The interpretation will remain valid so long as the paraxial assumption holds.

To begin, we note that the OPL in the context of a propagating wave is defined (see, for example, Saleh \cite{Saleh:91}) as
\begin{align}
    \Phi(\vec{x},z)=z+\phi(\vec{x},z)/(2k_0)
    \label{eqn:OPL}
\end{align}
that is, as the phase argument of the wave normalized by the wavenumber.  The quantity $\phi/(2k_0)$ can be viewed as a perturbation to the optical path resulting from refractive index fluctuations, diffraction, and, in this context, from polarization angle gradients as well.  
\par Consider now the expression for total path length $L(\vec{x}_z)$ in Lagrangian coordinates. For a differential length element parameterized by propagation distance, $d\ell(z)$
\begin{align}
L(\vec{x}_z)&=\int_0^z d\ell(\zeta)\nonumber \\
&=\int_0^z (d\zeta^2+dx^2_\zeta+dy^2_\zeta)^{1/2}=\int_0^z \left(1+\left(\frac{dx_\zeta}{d\zeta}\right)^2+\left(\frac{dy_\zeta}{d\zeta}\right)^2\right)^{1/2}d\zeta\nonumber \\
&\approx \int_0^z d\zeta+\frac{1}{2}\int_0^z \left(\frac{dx_\zeta}{d\zeta}\right)^2 d\zeta+\frac{1}{2}\int_0^z \left(\frac{dy_\zeta}{d\zeta}\right)^2 d\zeta
\end{align}
where in the final step we have made use of the fact that for $\epsilon_1,\epsilon_2\ll 1$, $(1+\epsilon_1+\epsilon_2)^{1/2}\approx 1+\frac{1}{2}\epsilon_1+\frac{1}{2}\epsilon_2$ which is equivalent to the paraxial assumption.  Equating this expression to the OPL given in Eqn. (\ref{eqn:OPL}) gives
\begin{align}
L(\vec{x}_z)=z+\frac{\phi(\vec{x}_z)}{2k_0}=\int_0^z d\zeta+\frac{1}{2}\int_0^z \left(\frac{dx_\zeta}{d\zeta}\right)^2 d\zeta+\frac{1}{2}\int_0^z \left(\frac{dy_\zeta}{d\zeta}\right)^2 d\zeta
\end{align}
so that
\begin{align}
\frac{\phi(\vec{x}_z)}{2k_0}&=\frac{1}{2}\int_0^z \left(\frac{dx_\zeta}{d\zeta}\right)^2 d\zeta+\frac{1}{2}\int_0^z \left(\frac{dy_\zeta}{d\zeta}\right)^2 d\zeta
\end{align}
Differentiating with respect to $z$ yields
\begin{align}
\frac{1}{k_0}\left(\frac{\partial\phi(\vec{x}_z)}{\partial x_z}\frac{\partial x_z}{\partial z}+\frac{\partial\phi(\vec{x}_z)}{\partial y_z}\frac{\partial y_z}{\partial z}\right)&= \left(\frac{dx_z}{dz}\right)^2+\left(\frac{dy_z}{dz}\right)^2.
\end{align}
which can be written
\begin{align}
\frac{1}{k_0}\nabla_{X_z}\phi(\vec{x}_z)\cdot \vec{v}(\vec{x}_z)&=\vec{v}(\vec{x}_z)\cdot \vec{v}(\vec{x}_z)
\label{eqn:velocity}
\end{align}
thus
\begin{align}
\vec{v}(\vec{x}_z)&=\frac{1}{k_0}\nabla_{X_z}\phi(\vec{x}_z).
\label{eqn:vel}
\end{align}
This demonstrates that the change in optical path length in the transverse plane per unit change in $\hat{z}$ (i.e. the transverse ``velocity'') is equal to the transverse phase gradient normalized by the wavenumber.  Eq. (\ref{eqn:vel}) was presented in \cite{Petruccelli:13} (Eq. 11 of that work) and is referred to appropriately as the ``flow vector of spectral density''. The connection between $\vec{v}(\vec{x}_z)$ and transverse coordinate changes was also leveraged in \cite{Nichols:19} although not explicitly derived as we have here.  The key assumption used to arrive at this result is that changes in the transverse direction are much smaller than changes in the direction of propagation and is tantamount to the same paraxial assumption we have used throughout our model development.

While the above analysis is performed for the coherent case, we have demonstrated that one can re-define the model quantities (amplitude, phase, and polarization angle) as appropriate averages in the partially coherent case. Thus, the above interpretation still holds and the transverse phase gradient can be interpreted as the average change in optical path in the transverse direction per unit change in the direction of propagation.

\section{Derivation of Equation (\ref{eqn:Jacobian}) \label{sec:Jacobian}}

The identity required of Eqn. (\ref{eqn:Jacobian}) requires relating the determinant of the Jacobian of the Lagrangian coordinate mappings $x_z(x_0,y_0),~y_z(x_0,y_0)$ to the divergence of the associated velocity field $u_z(x_z,y_z),~v_z(x_z,y_z)$.  Begin with the determinant
\begin{align}
    \det(J_{\vec{x}_0}(\vec{x}_z))&=\frac{dx_z}{dx_0}\frac{dy_z}{dy_0}-\frac{dy_z}{dx_0}\frac{dx_z}{dy_0},
\end{align}
take the derivative with respect to $z$ and recognize that $u_z\equiv dx_z/dz,~v_z\equiv dy_z/dz$.  Then  
\begin{align}
    \frac{d}{dz}\det(J_{\vec{x}_0}(\vec{x}_z))&=\frac{du_z}{dx_0}\frac{dy_z}{dy_0}+\frac{dv_z}{dy_0}\frac{dx_z}{dx_0}-\frac{dv_z}{dx_0}\frac{dx_z}{dy_0}-\frac{du_z}{dy_0}\frac{dy_z}{dx_0}
    \label{eqn:timederiv}
\end{align}
Both $u_z(x_z,y_z),~v_z(x_z,y_z)$ are functions of the Lagrangian coordinates, hence we can apply the chain rule to obtain
\begin{align}
\frac{du_z}{dx_0}&=\frac{\partial u_z}{\partial x_z}\frac{\partial x_z}{\partial x_0}+\frac{\partial u_z}{\partial y_z}\frac{\partial y_z}{dx_0}\nonumber \\
\frac{d u_z}{d y_0}&=\frac{\partial u_z}{\partial x_z}\frac{\partial x_z}{dy_0}+\frac{du_z}{\partial y_z}\frac{\partial y_z}{\partial y_0}\nonumber \\
\frac{d v_z}{d x_0}&=\frac{\partial v_z}{\partial x_z}\frac{\partial x_z}{\partial x_0}+\frac{\partial v_z}{\partial y_z}\frac{\partial y_z}{\partial x_0}\nonumber \\
\frac{d v_z}{d y_0}&=\frac{\partial v_z}{dx_z}\frac{\partial x_z}{\partial y_0}+\frac{\partial v_z}{\partial y_z}\frac{\partial y_z}{\partial y_0}.
\label{eqn:derivs}
\end{align}
Substituting (\ref{eqn:derivs}) into (\ref{eqn:timederiv}) and simplifying yields the ordinary differential equation
\begin{align}
\frac{d\det(J_{\vec{x}_0}(\vec{x}_z))}{dz}&=\left(\nabla_{X_s}\cdot\vec{v}(\vec{x}_s\right)\det(J_{\vec{x}_0}(\vec{x}_z)),
\end{align}
with solution
\begin{align}
\det\left(J_{\vec{x}_0}(\vec{x}_z)\right)&=\exp\left(\int_{s=0}^z \nabla_{X_s}\cdot\vec{v}(\vec{x}_s)ds\right)
\end{align}
This relationship allows (\ref{eqn:divform}) to be written as (\ref{eqn:Jacobian}) thus completing the derivation.

\section{Derivation of Equation (\ref{eqn:polarcons}) \label{sec:polar}}

To derive Equation (\ref{eqn:polarcons}) we begin with Eqn. (\ref{eqn:veceikonal})
\begin{align}
\partial_z\left(\rho_0\vec{\Omega}\right)+\nabla_X\cdot\left[\rho_0\vec{v}\otimes\vec{\Omega}+\rho_0\vec{\Omega}\otimes \vec{v}\right]= 0
\end{align}
Using two applications of the vector identity $\nabla_X\cdot\left(\vec{B}\otimes\vec{A}\right)=\vec{A}\left(\nabla_X\cdot\vec{B}\right)+\left(\vec{B}\cdot\nabla_X\right)\vec{A}$ and grouping terms we have

\begin{align}
\rho_0[\partial_z\vec{\Omega}+(\vec{v}\cdot\nabla_X)\vec{\Omega}]&+
\vec{\Omega}\left[\cancel{\partial_z(\rho_0)+\nabla_X\cdot\rho_0\vec{v}}\right]+\vec{v}\cancel{\left(\nabla_X\cdot [\rho_o\vec{\Omega}]\right)}+\left(\rho_0\vec{\Omega}\cdot\nabla_X\right)\vec{v}=0
\label{eqn:expand}
\end{align}
where Eqn. (\ref{eqn:divgrad}) has allowed us to cancel the second to last term while continuity of the first Stokes parameter (i.e., the intensity) cancels the third term.  Noting the first term in brackets is the total derivative of $\vec{\Omega}$, allows us to write Eqn. (\ref{eqn:expand}) in Eulerian coordinates as
\begin{align}
\frac{D\vec{\Omega}}{Dz}+\left(\vec{\Omega}\cdot\nabla_X\right)\vec{v}=0
\label{eqn:EulerForm}
\end{align} 
However, further simplification is possible by re-writing (\ref{eqn:expand}) as
\begin{align}
\partial_z\vec{\Omega}+\left\{\left(\vec{v}\cdot\nabla_X\right)\Omega+\left(\vec{\Omega}\cdot\nabla_X\right)\vec{v}\right\}&=0\nonumber \\
\partial_z\vec{\Omega}+\nabla_X\left(\vec{\Omega}\cdot\vec{v}\right)&=0.
\label{eqn:contAgain}
\end{align}
where we have leveraged the identity $\nabla_X \left(\vec{A}\cdot\vec{B}\right)=(\vec{A}\cdot\nabla_X)\vec{B}+(\vec{B}\cdot\nabla_X)\vec{A}+\vec{A}\times(\nabla_X\times\vec{B})+\vec{B}\times(\nabla_X\times\vec{A})$ and note that both $\vec{\Omega}$ and $\vec{v}$ are expressible as gradients of scalars, hence the cross-product terms vanish. 

Recalling that $\vec{\Omega}\equiv k_0^{-1} \nabla_X\gamma$ we can write
\begin{align}
k_0^{-1} \left[\partial_z(\nabla_X\gamma)+\nabla_X(\nabla_X\gamma\cdot\vec{v})\right]&=0\nonumber \\
=k_0^{-1}\nabla_X\left[\partial_z\gamma+\nabla_X\gamma\cdot\vec{v}\right]&=0\nonumber \\
=k_0^{-1}\nabla_X\frac{D\gamma}{Dz}&=0.
\end{align}
If we now make the switch to Lagrangian coordinates, the total derivatives become ordinary derivatives 
%and the second term in (\ref{eqn:EulerForm}) vanishes (gradient with respect to $\vec{x}_z$ of $d\vec{x}_z/dz$ is zero) leaving
\begin{align}
\frac{d\vec{\Omega}(\vec{x}_z,\omega)}{dz}=0.
\end{align}
which is Eqn. (\ref{eqn:polarcons}).

\bibliographystyle{plain}
\bibliography{main}
%\\*[0.5in]
%\begin{align}
%i2k_0\partial_z u(\vec{x},z)+\nabla_X^2 u(\vec{x},z)=0
%\end{align}
%solved by
%\begin{align}
%u(\vec{x},z)=\frac{k}{2\pi i z}\int\int u(\vec{x},0)e^{\frac{ik(\vec{x}-\vec{x}_0)^2}{2z}}d\vec{x}_0
%\end{align}
%if
%\begin{align}
%u(\vec{x},0)=Ae^{-\frac{x_0^2 + y_0^2}{2\sigma^2}}
%\end{align}
%then
%\begin{align}
%\rho(\vec{x},z)=\frac{A^2k^2\sigma^4}{z^2+k^2\sigma^2}e^{-\frac{k^2(x^2 + y^2)\sigma^2}{z^2+k^2\sigma^4}}
%\end{align}
\end{document}